\documentclass[usenatbib,usegraphicx,usepsfrag,twocolumn]{mn2e}

\newcommand{\be}{\begin{equation}}
\newcommand{\e}{\end{equation}}
\newcommand{\bear}{\begin{eqnarray}}
\newcommand{\ear}{\end{eqnarray}}

\def\I{I_{\nu}}

\def\pasa{PASA}
\def\mnras{MNRAS}
\def\aaps{A\&AS}
\def\apj{Ap.J}
\def\apjs{Ap.JS}

\def\aap{A\&A}
\def\araa{ARA\&A}
\def\u{{\bf U}} 
\def\th{\vec{\theta}}
\def\V{\mathcal{V}}

\usepackage{psfrag}
\def\del{\partial}
\begin{document}
\date {}
\title[Foreground for EoR : $150 \, {\rm MHz}$ GMRT observation]
{Characterizing Foreground for redshifted $21 \,{\rm cm}$ radiation:
  $150 \, {\rm MHz}$ GMRT observations}

\author[A. Ghosh,J. Prasad, S. Bharadwaj, S. S. Ali and
  J. N. Chengalur]{Abhik Ghosh$^{1}$\thanks{E-mail:
    abhik@phy.iitkgp.ernet.in}, Jayanti
  Prasad$^{2}$\thanks{E-mail: jayanti@iucaa.ernet.in }, Somnath
  Bharadwaj$^{1}$\thanks{Email:somnath@phy.iitkgp.ernet.in}, Sk. Saiyad
  Ali$^{3}$\thanks{Email:saiyad@phys.jdvu.ac.in} and \and Jayaram
  N. Chengalur$^{4}$\thanks{Email:chengalu@ncra.tifr.res.in} \\ $^{1}$
  Department of Physics and Meteorology \& Centre for Theoretical
  Studies , IIT Kharagpur, 721 302 , India \\$^{2}$ IUCAA, Post Bag 4,
  Ganeshkhind, Pune University Campus, Pune 411 007, India \\$^{3}$
  Department of Physics,Jadavpur University, Kolkata 700032, India.
  \\$^4$ National Centre for Radio Astrophysics, TIFR, Post Bag 3,
  Ganeshkhind, Pune 411 007, India}

\maketitle
   
\begin{abstract}
Foreground removal is a major challenge for detecting the redshifted
$21$-cm neutral hydrogen (HI) signal from the Epoch of Reionization
(EoR). We have used $ 150 \, {\rm MHz}$ GMRT observations to
characterize the statistical properties of the foregrounds in four
different fields of view.  The measured multi-frequency angular power
spectrum $C_{\ell}(\Delta \nu)$ is found to have values in the range
$10^4 \ {\rm mK^2}$ to $2 \times 10^4 \ {\rm mK}^2$ across $700 \le
\ell \le 2 \times 10^4$ and $\Delta \nu \le 2.5 \ {\rm MHz}$, which is
consistent with model predictions where point sources are the most
dominant foreground component.  The measured $C_{\ell}(\Delta \nu)$
does not show a smooth $\Delta \nu$ dependence, which poses a severe
difficulty for foreground removal using polynomial fitting.

The observational data was used to assess point source
subtraction. Considering the brightest source $(\sim 1 \ {\rm Jy})$ in
each field, we find that the residual artifacts are less than $1.5 \%$
in the most sensitive field (FIELD I). Considering all the sources in
the fields, we find that the bulk of the image is free of artifacts,
the artifacts being localized to the vicinity of the brightest
sources. We have used FIELD I, which has a rms noise of $1.3 \ {\rm
  mJy/Beam}$, to study the properties of the radio source population
to a limiting flux of $9 \ {\rm mJy}$. The differential source count
is well fitted with a single power law of slope $-1.6$.  We find there
is no evidence for flattening of the source counts towards lower flux
densities which suggests that source population is dominated by the
classical radio-loud Active Galactic Nucleus (AGN).

The diffuse Galactic emission is revealed after the point sources are
subtracted out from FIELD I . We find $C_{\ell} \propto \ell^{-2.34}$
for $253 \le \ell \le 800$ which is characteristic of the Galactic
synchrotron radiation measured at higher frequencies and larger
angular scales.  We estimate the fluctuations in the Galactic
synchrotron emission to be $ \sqrt{\ell(\ell+1)C_{\ell}/2\pi} \simeq
10 \, {\rm K}$ at $\ell=800$ ($\theta > 10'$). The measured $C_{\ell}$
is dominated by the residual point sources and artifacts at smaller
angular scales where $C_{\ell} \sim 10^3 \ {\rm mK}^2$ for $\ell >
800$.
\end{abstract} 


\section{Introduction}

Observations of the high-redshift Universe with the 21-cm hyperfine
line emitted by neutral hydrogen gas (HI) during the Epoch of
Reionization (EoR) contains a wealth of cosmological and astrophysical
informations (see \citealt{furla}; \citealt{morales10} for recent
reviews). Analysis of quasar absorption spectra \citep{becker,fan} and
the CMBR \citep{komatsu} together indicate that reionization probably
started around a redshift of $15$ and lasted until a redshift of
$6$. The Giant Metrewave Radio Telescope
(GMRT \footnote{http://www.gmrt.ncra.tifr.res.in}; \citealt{swarup})
is currently functioning at several bands in the frequency range of
$150 \, {\rm MHz}$ to $1420 \, {\rm MHz}$ and can potentially detect
the 21 cm signal at high redshifts.  Several upcoming low-frequency
telescopes such as LOFAR\footnote{http://www.lofar.org/},
MWA\footnote{http://www.mwatelescope.org/} and
PAPER\footnote{http://astro.berkeley.edu/\textasciitilde dbacker/eor/}
are also targeting a detection of the high redshift 21-cm signal.
LOFAR is already operational and will soon start to collect data for
the EoR project.

In this paper we have carried out GMRT observations to characterize
the statistical properties of the background radiation at $150 \, {\rm
  MHz}$.  These observations cover a frequency band of $\sim 5 \ {\rm
  MHz}$ with a $\sim 100 \ {\rm kHz}$ resolution, and cover the
angular scales $\sim 30'$ to $\sim 30''$.  The 21 cm radiation is
expected to have an rms brightness temperature of a few mK on angular
scales of a few arc minute \citep{zald}. This signal, is however,
buried in the emission from other astrophysical sources which are
collectively referred to as foregrounds.  Foreground models
\citep{ali} predict that they are dominated by the extragalactic radio
sources and the diffuse synchrotron radiation from our own Galaxy
which makes a smaller contribution. Analytic estimates of the HI 21-cm
signal indicate that the predicted signal decorrelates rapidly with
increasing $\Delta\nu$ and the signal falls off $90\%$ or more at
$\Delta\nu \ge 0.5 \ {\rm MHz}$ \citep{BS1,BA5}. The foreground from
different astrophysical sources are expected to be correlated over a
large frequency frequency separation. This property of the signal
holds the promise of allowing us to separate the signal from the
foregrounds \citep{ghosh2}. In this paper we have characterized the
foreground properties across a frequency separation $\Delta \nu$ of
$2.5 \ {\rm MHz}$, the signal may be safely assumed to have
decorrelated well within this frequency range.

There currently exist several surveys which cover large regions of the
sky at around $150 \,{\rm MHz}$ like the 3CR survey
\citep{edge,Bennett}, the 6C survey \citep*{hales,waldram} and the 7C
survey \citep{hales07}.  These surveys have a relatively poor angular
resolution and a limiting flux density of $\sim 100\, {\rm mJy}$.
\citet{dmat1} have used the 6C survey, the 3CR survey and the 3 CRR
catalogue \citep{laing} to estimate the extragalactic point source
contribution at $150 \, {\rm MHz}$.  The GMRT, which is currently the
most sensitive telescope at $150 \ {\rm MHz}$, offers the possibility
to perform deeper surveys with a higher angular resolution.  A few
single pointing $150 \, {\rm MHz}$ surveys
\citep{sirothia,ishwara10,ishwara,intema} have been performed with an
rms noise of $\sim 1 \ {\rm mJy/Beam}$ at a resolution of $20''-30''$.
The TGSS \footnote{http://tgss.ncra.tifr.res.in/} is currently
underway to survey the sky north of the declination $-30^{\circ}$ with
an rms noise of $7-9 \ {\rm mJy/Beam}$ at an angular resolution of
$20''$.

Diffuse synchrotron radiation produced by cosmic ray electrons
propagating in the magnetic field of our Galaxy \citep{GS69} is a
major foreground component. This component, which is associated with
our Galactic disk, is particularly strong in the plane of our
Galaxy. Our target fields were selected well off the Galactic plane
with $b>10^{\circ}$ where we expect a relatively lower contribution
from the Galactic synchrotron radiation.

Radio surveys at 408 MHz \citep{haslam}, 1.42 GHz \citep{R82,RR88} and
2.326 GHz \citep{JBN98} have measured the diffuse Galactic synchrotron
radiation at angular scales larger than $\sim 1^{\circ}$ where it is
the most dominant contribution, exceeding the point sources.
\citet{platania1} have analyzed these observations to show that the
Galactic synchrotron emission has a steep spectral index of ${\alpha}
\approx 2.8$.  The angular power spectrum $C_{\ell}$ has been
measured at $2.3 \,{\rm GHz}$ \citep{giardino01} and $2.4 \,{\rm GHz}$
\citep{giardino02} where they find $C_{\ell} \sim \ell^{-2.4}$ at
angular multipoles $\ell < 250$. WMAP data shows $C_{\ell} \sim
\ell^{-2}$ at $\ell \le 200$ \citep{bennett03} which is slightly
flatter compared to the findings at lower frequencies.  It is relevant
to note that all of these results are restricted to angular scales
greater than $43'$ $(\ell < 250)$, little is known (except
\citealt{bernardi09} discussed later) about the structure of the
Galactic synchrotron radiation at the sub-degree angular scales probed
in this paper.

\citet{ali} have carried out GMRT observations to characterize the
foregrounds on sub-degree angular scales at $150 \ {\rm MHz}$. They
have used the measured visibilities to directly determine the
multi-frequency angular power spectrum $C_{\ell}(\Delta \nu)$
(MAPS; \citet{kanan}) which characterizes the statistical properties of
the fluctuations in the background radiation jointly as a function of
the angular multipole $\ell$ and frequency separation $\Delta
\nu$. They find that the measured $C_{\ell}(\Delta \nu)$ has a value
of around $10^4 \ {\rm mK}^2$ which is seven orders of magnitude
larger than the expected HI signal. The measured $C_{\ell}(\Delta
\nu)$ was also found to exhibit relatively large ($\sim 10 - 20 \%$)
fluctuations in $\Delta \nu$ which pose a severe problem for
foreground removal.  Further, it was attempted to remove the
foregrounds by subtracting out all the identifiable point sources
above $8 \, {\rm mJy}$. The resulting $C_{\ell}$, however, only
dropped to $\sim 3$ times the original value due to residual
artifacts.
 
\citet{bernardi09} have analyzed $150 \, {\rm MHz}$ WSRT observations
near the Galactic plane ($(l=137^{\circ}, b=+8^{\circ}$) to
characterize the statistical properties of the diffuse Galactic
emission (after removing the point sources) at angular multipoles
$\ell < 3000$ ($\theta > 3.6'$).  They find that the measured total
intensity angular power spectrum shows a power law behaviour $C_{\ell}
\propto \ell^{-2.2}$ for $\ell \le 900$.  They also measured the
polarization angular power spectrum for which they find $C_{\ell}
\propto \ell^{-1.65}$ for $\ell \le 2700$. Their observations indicate
fluctuations of the order of $\sim 5.7 \ {\rm K}$ and $3.3 \ {\rm K}$
on $5'$ angular scales for the total intensity and the polarized
diffuse Galactic emission respectively.  \citet{pen09} have carried
out $150 \, {\rm MHz}$ GMRT observations at a high Galactic latitude
to place an upper limit of $\sqrt{\ell^2C_{\ell}/2\pi} < 3 {\rm K}$ on
the polarized foregrounds at $\ell < 1000$.

There has been a considerable amount of work towards simulating the
foregrounds \citep{jelic,sun,bowman09,gleser,harker,Liu1,liu09b,nada}
in order to develop algorithms to subtract the foregrounds and detect
the redshifted 21-cm signal. These simulations require guidance from
observational data, and it is crucial to accurately characterize the
foregrounds in order to have realistic simulations for future
observations.

One of the challenges for the detection of this cosmological 21-cm
signal is the accuracy of the foreground source removal. Our current
paper aims at characterizing the foreground at arcminute angular
scales which will be essential for extracting the 21-cm signal from
the data. In this paper we study the radio source population at $150
\, {\rm MHz}$, which is still less explored, to determine the
differential source count which plays a very important role in
predicting the angular power spectrum of point sources
\citep{ali,ghosh1}. We find that the point sources are the most
dominating foreground component at the angular scales of our
analysis. The accuracy of point source removal is one of the principal
challenges for the detection of the 21-cm signal. We investigate in
detail the extragalactic point source contamination in our observed
fields and study how accurately the bright point sources can be
removed from our images. We also estimate the level of residual
contamination in our images. It is expected that the diffuse Galactic
emission will be revealed if the point sources are individually
modelled and removed with high level of precession. To our knowledge,
the present work is only the second time that the fluctuations in the
Galactic diffuse emission have been detected at around $10$ arcminute
angular scales at a frequency of $150 \, {\rm MHz}$ .  We also
complement this with a measurement of the angular power spectrum of
the diffuse Galactic foreground emission in the $\ell$ range $200$ to
$800$.

In this paper we have analyzed GMRT observations to characterize the
foregrounds in four different fields at $150 \, {\rm MHz}$ which
corresponds to the HI signal from $z = 8.3$. We note that unless
otherwise stated we use the cosmological parameters
$(\Omega_{m0},\Omega_{\Lambda0}, \Omega_b h^2, h,
n_s,\sigma_8)=(0.3,0.7, 0.024,0.7,1.0,1.0)$ in our analysis. We now
present a brief outline of the paper. Section 2 describes the
observations and data analysis.  We have used the multi-frequency
angular power spectrum $C_{\ell}(\Delta \nu)$ to quantify the
statistical properties of the background radiation in the observed
fields. Section 3 presents the technique that we have used to estimate
$C_{\ell}(\Delta \nu)$ directly from the measured visibilities.  We
show the the measured $C_{\ell}(\Delta \nu)$, and discuss its
properties in Section 4. In this section we have also compared the
measured $C_{\ell}(\Delta \nu)$ with foreground model predictions.  In
Section 5 we consider the extragalactic point sources which forms the
most dominant foreground component in our observations. In order to
assess how well we can subtract out the point sources, we consider
point source subtraction for the brightest source in each field, as
well as subtracting out all the identifiable sources.  We use the most
sensitive field (FIELD I) in our observation to study the nature of
the radio source population at $150 \, {\rm MHz}$, and we also provide
a source catalogue for this field in the online version of this paper
(Appendix \ref{catal}).  In Section 6 we have used the residual data
of FIELD I, after point source removal, to study the diffuse Galactic
synchrotron radiation. We have summarized the results of the entire
paper in Section 7.
 
\section{GMRT Observations and data analysis}

The GMRT has a hybrid configuration (\citealt{swarup}), each of
diameter $\rm 45\, m$, where $14$ of the $30$ antennas are randomly
distributed in a Central Square which is approximately $\rm 1.1 \, km
\times 1.1\, km$ in extent.  The rest of the antennas lie along three
$\sim 14 \, {\rm km}$ long arms in an approximately `Y'
configuration. The shortest antenna separation (baseline) is around
$60 \, \rm m$ including projection effects while the largest
separation is around $26 \, \rm km$.  The hybrid configuration of the
GMRT gives reasonably good sensitivity to probe both compact and
extended sources. Table \ref{tab:obs_sum} summarizes our
observations including the calibrators used for each field. For all
observations, visibilities were recorded for two circular
polarisations (RR and LL) with $128$ frequency channels and $16 \,
{\rm s}$ integration time.

We have observed FIELD I in GTAC (GMRT Time Allocation Committee)
cycle 15 in January 2008, whereas FIELD II and III were observed in
cycle 17 during February 2010. Our target fields were selected at high
Galactic latitudes ($b>10^{\circ}$) which were up at night time during
the GTAC cycle 15 and 17, and which contain relatively few bright
sources ($\ge 0.3 \ {\rm Jy}$) in the 1400 MHz NVSS survey. We have
selected such fields where the 408 MHz Haslam map sky temperature is
relatively low (in the range of 30 K to 40 K) and there is no
significant structure visible at the angular resolution $(\sim
0.85^{\circ}$) of Haslam map. FIELD IV had been observed in cycle 8
(June 2005) and the $150 \ {\rm MHz}$ foregrounds in this field have
been analyzed and published earlier \citep{ali}. We have used the
calibrated data set of \citet{ali}, the foreground characterisation
has, however, been redone using an improved technique \citep{ghosh2}.

The observations were largely carried out at night to minimise the
Radio Frequency Interference (RFI) from man-made sources.  Further,
the ionosphere is considerably more stable at night, and the
scintillation due to ionospheric oscillations is expected to go down.
The duration of our observations ($\sim$ 4 - 10 hrs.) was chosen so as
to achieve a reasonably high signal to noise ratio, and also to
have a reasonably large number of visibilities adequate for a
statistical analysis of the foregrounds.

We have followed the standard procedure of interferometric
observations where we observe a flux calibrator of known strength and
a phase calibrator whose structure is known but whose strength varies
from epoch to epoch. The flux calibrator, which is observed at the
start of the observation, sets the amplitude scale of the gains and
this is used to set the amplitude of the phase calibrator. Phase
calibrators were chosen near the target fields, and these were
observed every half an hour so as to correct for temporal variations
in the system gain. We have used the observations of the phase
calibrators to determine the complex gains of the antennas, and these
gains were interpolated to the observations of the target fields.

\begin{table*}
\caption{Observation summary} 
\vspace{.2in}
\label{tab:obs_sum}
\begin{tabular}{|l|c|c||c||c|}
\hline
 & FIELD I & FIELD II & FIELD III& FIELD IV\\ 
\hline
Date & 2008 Jan 08 & 2010 Feb 07&  2010 Feb 08 &  2005 June 15 \\
\hline
Working antennas & 28 & 28 & 26 & 23 \\
\hline
Central Frequency & 153 MHz & 148 MHz & 148 MHz& 153 MHz \\
\hline
Channel width & 62.5 kHz & 125.0 kHz & 125.0 kHz& 62.5 kHz \\
\hline
Bandwidth & 3.75 MHz & 8.75 MHz & 10.0 MHz& 4.37 MHz \\
\hline
Total observation time &  11 hrs &  6 hrs &  6 hrs&  13 hrs\\
\hline
Flux calibrator& 3C147 & 3C147 & 3C286& 3C48\\
Observation time &  1.2 hrs&  1.75 hrs&  0.5 hrs&  2 hrs\\
Flux density (Perley-Taylor 99) & $\sim$ 72.5 Jy& $\sim$ 72.6 Jy& $\sim$ 33.41 Jy&$\sim$ 62.5 Jy\\
\hline
Phase calibrator& 3C147 &3C147  &1459+716 &3C48 \\
Observation time & - & - & 1.6 hrs& -\\
Flux density & - & - & $\sim$ 38.7 Jy& -\\
\hline
Target field  $(\alpha,\delta)_{2000}$ &  ($05^h30^m00^s$,&
($06^h00^m00^s$,& ($12^h36^m49^s$,& ($1^h36^m48^s$,\\ 
 & $+60^{\circ}00^{'}00^{''}$) & $  +62^{\circ}12^{'}58^{''} $) & 
$+62^{\circ}12^{'}58^{''}\, $) & $ +41^{\circ}24^{'}23^{''} $)\\
 Galactic coordinates $(l,b)$ & $151.80^{\circ}, 13.89^{\circ}$ &
$ 151.49^{\circ}, 18.12^{\circ}$ & $125.89^{\circ}, 54.83^{\circ}$ 
&$132.00^{\circ},  20.67^{\circ}$\\
Sky Temp. (Haslam Map, 408 MHz) & $\sim 40\, {\rm K}$&$\sim 30\, {\rm K}$&$\sim 20\, {\rm K}$ &$\sim 30\, {\rm K}$\\
Observation time &  9.8 hrs& 4.25 hrs & 3.9 hrs &  11 hrs\\ 
\hline
\hline
\end{tabular}
\end{table*}
\vspace{.2in}

\begin{table*}
\caption{Wide field Imaging summary}
\vspace{.2in}
\label{tab:img_sum}
\begin{tabular}{|c|c|c||c||c|}
\hline
 & FIELD I & FIELD II & FIELD III& FIELD IV\\ 
\hline
Image size &$5700 \times 5700$& $5500 \times 5500$ & $5500 \times 5500$ & $5500 \times 5500$ \\
\hline
Pixel size &$3.22'' \times 3.22''$  &$3.32'' \times 3.32''$  &$3.27'' \times 3.27''$  & $3.40'' \times 3.40''$ \\
\hline
Number of facets & 121 with 2 overlap  & 109 with 2 overlap &109 with 2 overlap & 109 with 2 overlap\\
\hline
Off-source Noise & $1.3 \, {\rm mJy/Beam}$ & $2.5 \, {\rm mJy/Beam}$ &$4.5 {\rm mJy/Beam}$ & $1.6 {\rm mJy/Beam}$\\
\hline
Peak/Noise & $700$ & $620$ &$422$ & $550$\\
\hline
Flux density (max., min.)& ($905 \,{\rm mJy/Beam}$, & ($1.55 \,{\rm Jy/Beam}$, & ($1.9 \,{\rm Jy/Beam}$, & ($900 \,{\rm mJy/Beam}$, \\
& $-14 \,{\rm mJy/Beam}$)&$-28 \,{\rm mJy/Beam}$)&$-45 \,{\rm mJy/Beam}$)&$-47 \,{\rm mJy/Beam}$)\\
\hline
Synthesized beam & $21'' \times 18''$, PA=$61^{\circ}$ & $30'' \times 22''$, PA=$-52^{\circ}$ &$33'' \times 20''$, PA=$53^{\circ}$ &$24'' \times 18''$, PA=$70^{\circ}$ \\
\hline
\hline
\end{tabular}
\end{table*}
\vspace{.2in}
The data for FIELDS I, II and III were reduced using FLAGCAL
(\citet{jayanti} which is a flagging and calibration software for
radio interferometric data. FIELD IV, however, was not analyzed with
FLAGCAL, it was flagged manually and then calibrated using the
standard task within Astronomical Image Processing Software
(AIPS). The details of the observation, data reduction and relevant
statistics for FIELD IV are discussed in \citet{ali}.

RFI is a major factor limiting the sensitivity of at low frequencies.
RFIs effectively increases the system noise and corrupts the
calibration solutions. It also restricts the available frequency
bandwidth. The effect is particularly strong at frequencies below 0.5
GHz. FLAGCAL identifies and removes bad visibilities by requiring that
good visibilities be continuous in time and frequency, computes
calibration solutions using known flux and phase calibrators and
interpolates these onto the target fields (see \citet{jayanti} for
more details ). The flagged and calibrated visibility data were used
to make continuum images using the standard tasks in the AIPS.

The large field of view (hereafter FoV; $\theta_{\rm
  FWHM}=3.8^{\circ}$) of the GMRT at $150 \, {\rm MHz}$ leads to
considerable error if the non-coplanar nature of the GMRT antenna
distribution is not taken into account. We use the three dimensional
(3D) imaging feature \citep{perley} in the AIPS task `IMAGR' in which
the entire field of view is divided into multiple sub-fields (facets),
each of which is imaged separately. Table \ref{tab:img_sum} contains a
summary of the imaging details with all the relevant parameters which
are mostly self-explanatory.

The presence of a large number of bright sources in the fields that we
have observed allows us to carry out self calibration to improve the
complex gains. This reduces the errors from temporal variations in the
system gain, and spatial and temporal variations in the ionospheric
properties.  For all the target fields, the data went through $3$
rounds of phase self calibration followed by one round of self
calibration for both amplitude and phase. The phase variations often
occur on timescales of as few minutes or less.  The time interval for
gain correction was chosen as $5, 3, 2$ and $5$ \,minutes for the
successive self calibration loops.  In successive self-calibration
loops, the gain solution converged rapidly with less than $0.1 \%$ bad
solutions in each loop. The final gain table was applied to all the
$128$ frequency channels.

\begin{figure*}
\includegraphics[width=75mm]{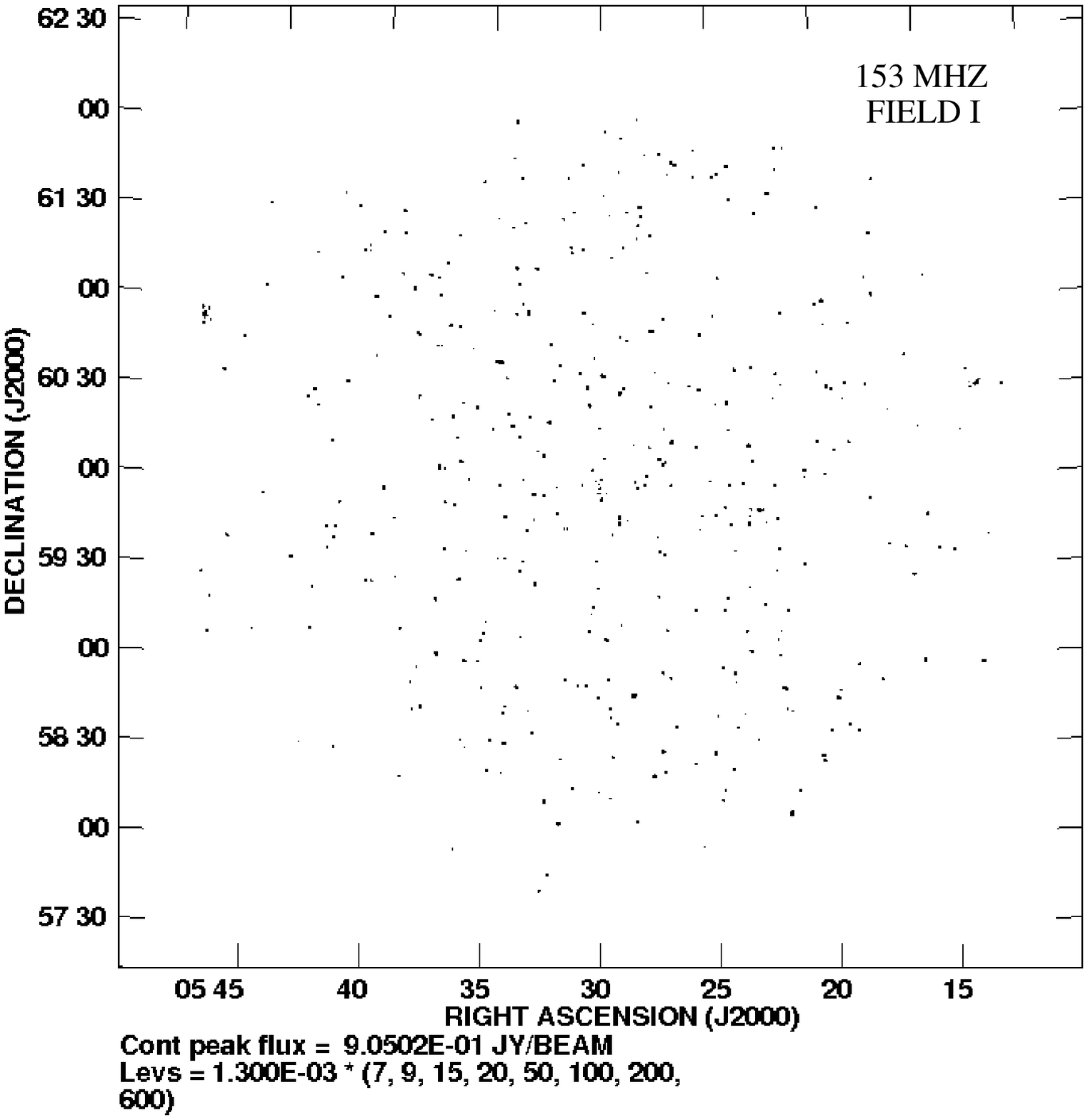}
\includegraphics[width=75mm]{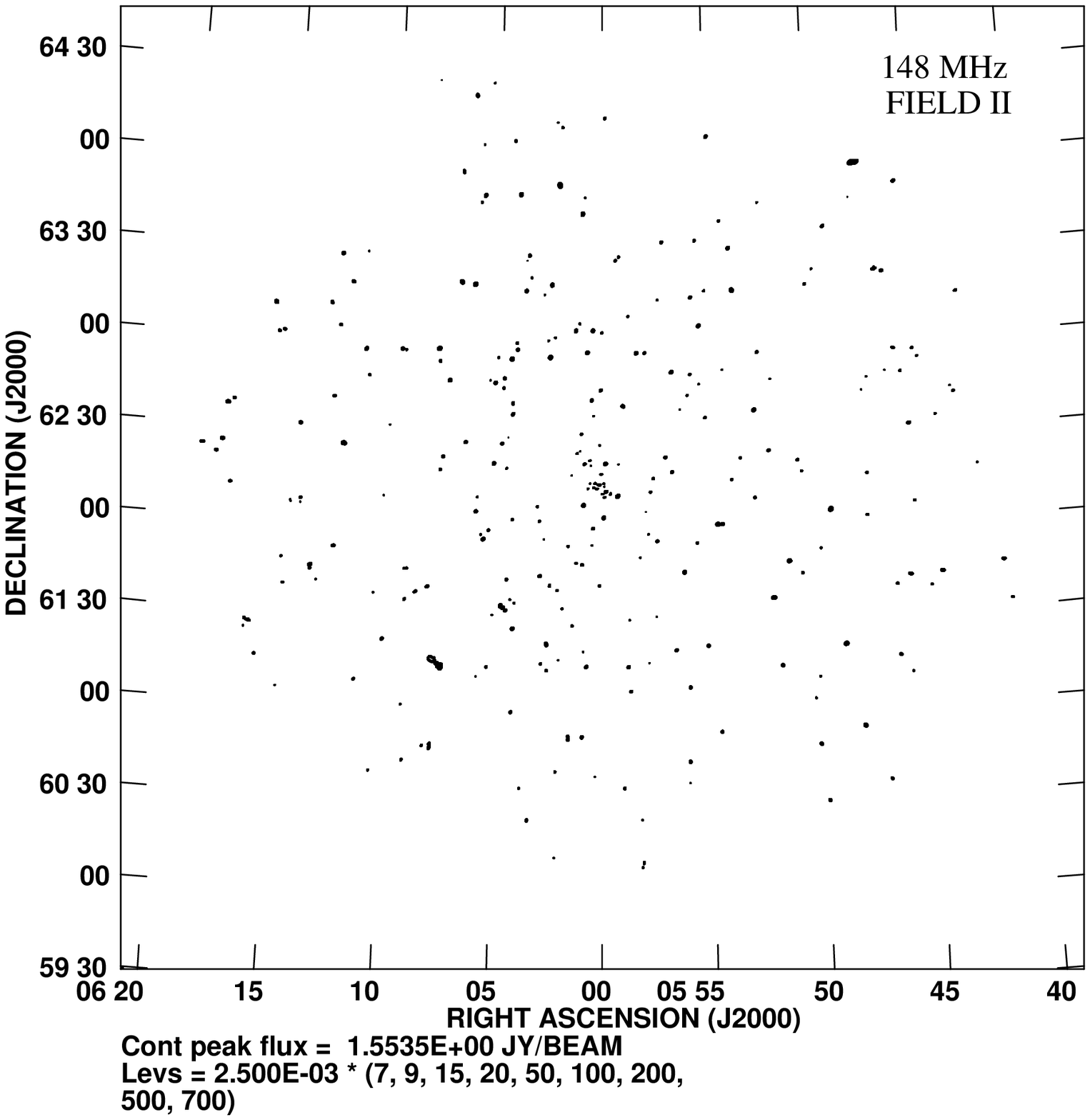}
\includegraphics[width=75mm]{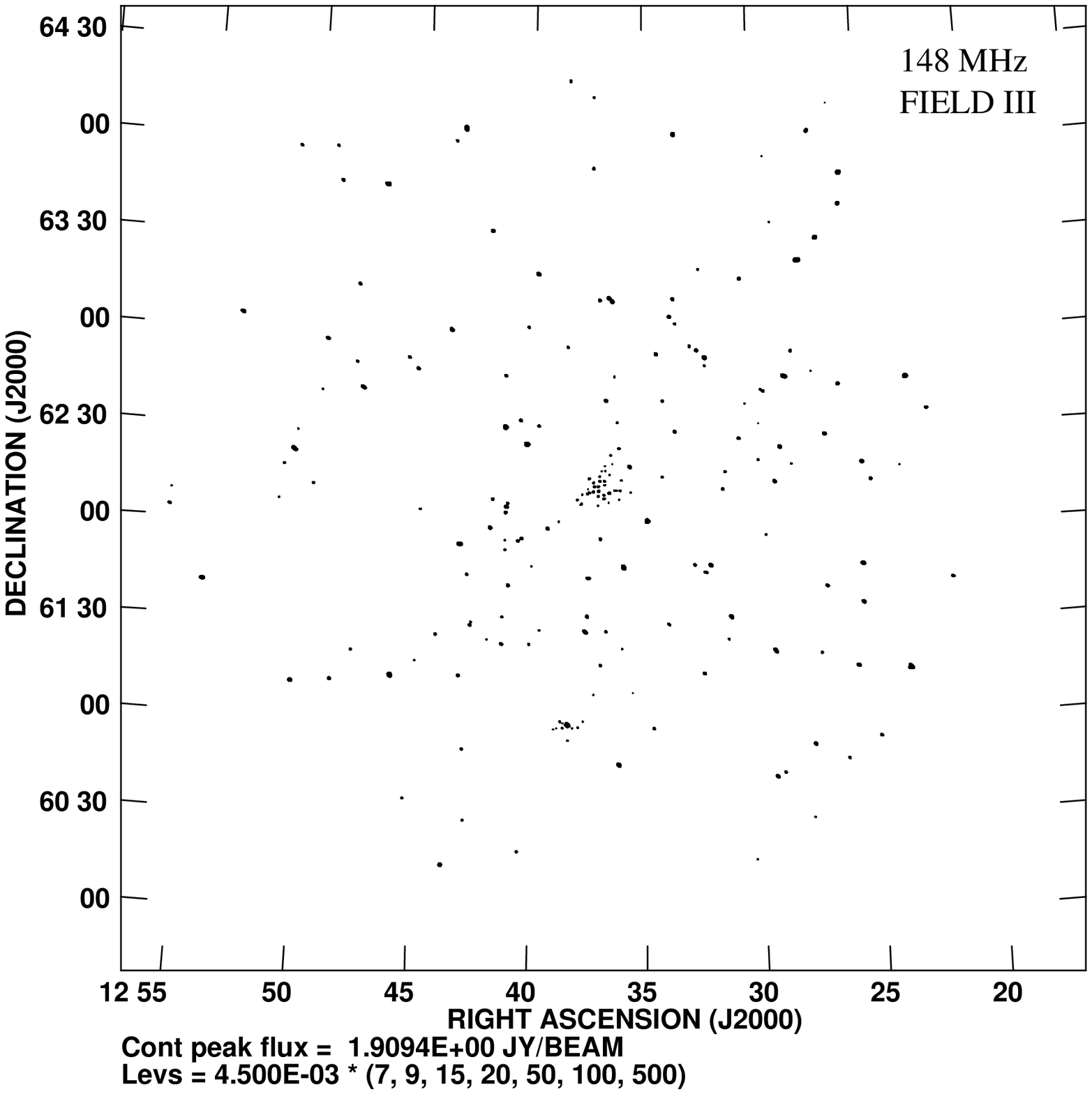}
\includegraphics[width=75mm]{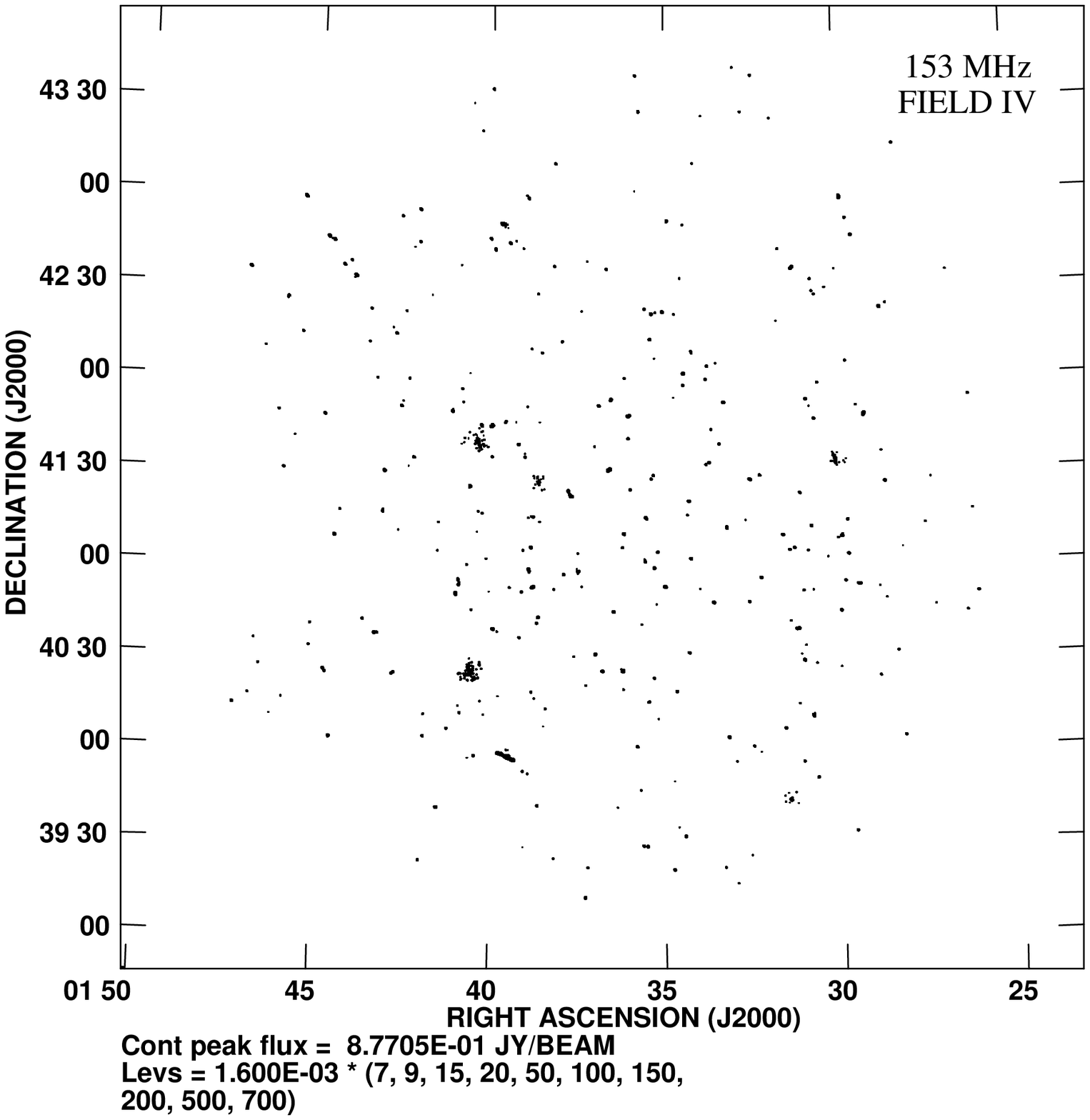}
\caption{The figure shows the continuum images of bandwidth $3.75 \,
  {\rm MHz}$, $8.75 \, {\rm MHz}$, $10.0 \, {\rm MHz}$ and $4.37 \,
  {\rm MHz}$ of FIELD I, FIELD II, FIELD III and FIELD IV
  respectively. The off source rms noise is around $1.3 {\rm
    mJy/Beam}$, $2.5 {\rm mJy/Beam}$, $4.5 {\rm mJy/Beam}$ and $1.6
  {\rm mJy/Beam}$ for FIELD I, FIELD II, FIELD III and FIELD IV
  respectively.}
\label{fig:fields}
\end{figure*}

The final continuum images for the four fields are shown in Figure
\ref{fig:fields}. The images all have a FoV of
$4.0^{\circ}\times4.0^{\circ}$.  To avoid bandwidth smearing in the
continuum image, we have collapsed adjacent channels within $\le 0.7\,
{\rm MHz}$ and separately combined the respective images.

The subsequent analysis was done using the final; calibrated
visibilities of the original 128 channel data. The final data contains
visibilities for $2$ circular polarisations. The visibilities from the
two circular polarisations were combined ($\V=[\V_{RR}+\V_{LL}]/2$)
for the subsequent analysis.

\section{ Estimation of angular power spectrum from visibility correlations}
\label{sec3}

In this paper we have used the multi-frequency angular power spectrum
$C_l(\Delta \nu)$ (\citealt{kanan}) to quantify the statistical
properties of the sky signal. This jointly characterizes the
dependence on angular scale $\ell^{-1}$ and frequency separation
$\Delta \nu$.  In this section we briefly present how we have
estimated $C_{\ell}(\Delta \nu)$ from the measured visibilities.

For a frequency $\nu$, the angular dependence of the brightness
temperature distribution on the sky $T({\bf \hat{n}},\nu)$ may be
expanded in spherical harmonics as

\begin{equation}
T({\bf \hat{n}},\nu)  = \sum_{\ell,m} \
a_{\ell m}(\nu) \ Y_{\ell m} ({\bf \hat{n}})\,.
\label{eq:t1}
\end{equation}

The multi-frequency angular power spectrum  is defined as

\begin{equation}
C_l(\Delta \nu) \equiv C_l(\nu, \nu + \Delta \nu)=\langle a_{lm}(\nu)
~ a^*_{lm}(\nu + \Delta \nu) \rangle \,.
\label{eq:c1}
\end{equation} 

where $\ell$ refers to the angular multipoles on the sky and
$\Delta\nu$ is the frequency separation.

It is possible to use the correlation between pairs of visibilities
$\V(\u,\nu)$ and $\V(\u+\Delta \u,\nu+ \Delta \nu)$ to estimate
$C_{\ell}(\Delta \nu)$ where $\ell= 2 \pi U$.

\begin{equation}
V_2(\u,\nu;\u + \Delta \u,\nu+\Delta \nu) \equiv \langle \V(\u,\nu) 
\V^*(\u+\Delta \u,\nu+ \Delta \nu)  \rangle
\label{eq:v2}
\end{equation}

The correlation of a visibility with itself is excluded to avoid a
positive noise bias in the estimator. \citet{ali} as well as
\citet{Dutta} contain detailed discussions of the estimator.

\begin{figure*}
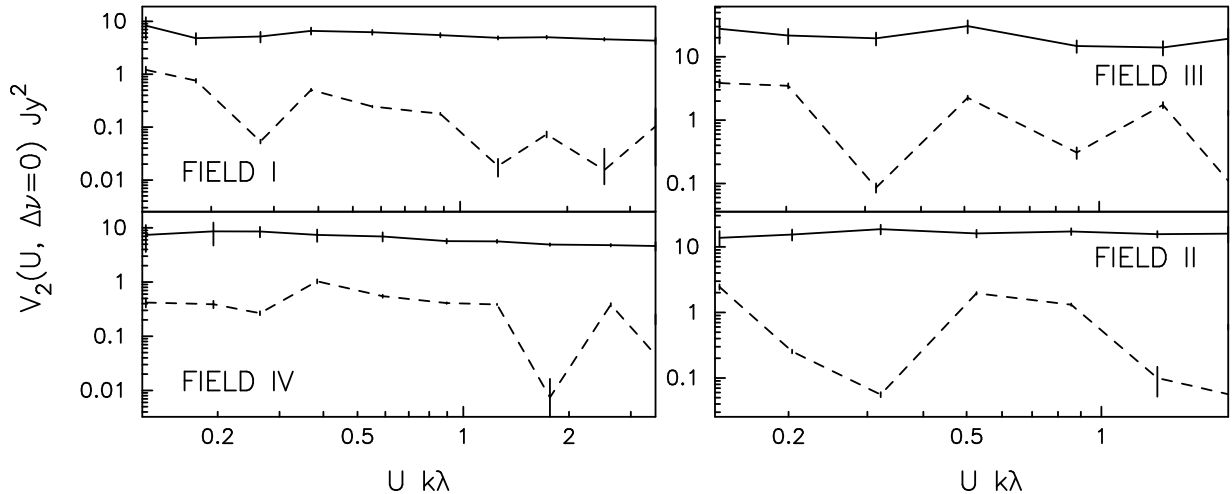

\includegraphics[width=65mm,angle=-90]{v2_14.ps}
\includegraphics[width=65mm,angle=-90]{v2_23.ps}
\caption{This shows the real (upper curve) and imaginary (lower curve)
  parts of the observed visibility correlation $V_2(U,0)$ as a
  function of baselines $U$. The $1 \sigma$ error-bars shown on the
  plot includes contribution from both cosmic variance and system
  noise.}
\label{fig:v2}
\end{figure*}

For the sky signal which is dominated by foregrounds the measured
$C_{\ell}(\Delta \nu)$, for a fixed $\ell$, is expected to vary
smoothly with $\Delta \nu$ and remain nearly constant over the
observational bandwidth. In a recent $610 \, {\rm MHz}$ observations
\citep{ghosh1} we find instead that in addition to a component that
exhibits a smooth $\Delta \nu$ dependence, the measured
$C_{\ell}(\Delta \nu)$ also has a component that oscillates as a
function of $\Delta \nu$ (around 1-4\%). We note that similar
oscillations, with considerably larger amplitudes, have been reported
in earlier GMRT observations at $153 \, {\rm MHz}$ \citep{ali}. In our
recent work \citep{ghosh2} we have noted that these oscillations could
arise from bright continuum sources located at large angular
separations from the phase center. The problem can be mitigated by
tapering the array's sky response with a frequency independent window
function $W(\vec{\theta})$ that falls off before the first null of the
primary beam (PB).  RFI sources, which are mostly located on the
ground, are picked up through the sidelobes of the primary beam.
Tapering the array's sky response is also expected to mitigate the RFI
contribution.

Close to the phase centre, the 150 MHz GMRT PB is reasonably well
modeled by a Gaussian $A(\th,\nu)=e^{-\theta^2/\theta_0^2}$ where the
parameter $\theta_0$ is related to the FWHM of the PB as $\theta_0
\approx 0.6 \times \theta_{\rm FWHM}$, and $\theta_0 =2.3^{\circ}$
($\theta_{\rm FWHM}=3.8^{\circ}$). We taper the array's sky response
by convolving the observed visibilities with a suitably chosen
function $\tilde{W}(\u)$.  The sky response of the convolved
visibilities $\V_c=\tilde{W} \otimes \V$ is modulated by the window
function $W(\th)$ which is the Fourier transform of $\tilde{W}(\u)$.
We have used a window function $W(\th)=e^{-\theta^2/\theta^2_w}$ with
$\theta_w < \theta_0$ to taper the sky response so that it falls off
well before the first null. We parametrise $\theta_w$ as $\theta_w=\rm
{f}\theta_0$ with $\rm{f} \leq 1$ where $\theta_0$ here refers to the
value at the fixed frequency $150 \, {\rm MHz}$.

We have used the convolved visibilities to determine the
two-visibility correlation which is defined as
 
\begin{equation}
 V_{2}(\u_i,\Delta\nu)=\V_c(\u_i,\nu) \V_c^{*}(\u_i,\nu+ \Delta \nu) \,.
\label{eq:2ve}
\end{equation}

 Here $\u_i$ refers to different points (labelled by $i$) on a grid in
 the $uv$ plane. We have used a grid of spacing $\Delta U_g=\sqrt{\ln
   2}/(\pi \theta_w)=0.265 \theta_w^{-1}$ which corresponds to half of
 the FWHM of $\tilde{W}(\u)$, and we have estimated
 $V_{2}(\u_i,\Delta\nu) $ at every grid point using all the baselines
 within a disk of radius $2 \, \Delta U_g$ centered on that grid
 point.  Using eq. (\ref{eq:2ve}), however, introduces an undesirable
 positive noise bias in the estimated $V_{2}(\u_i,\Delta\nu)$
 (e.g. \citealt{Begum}). It is possible to avoid the noise bias if we
 use the estimator

\begin{eqnarray}
V_{2}(\u_i,\Delta\nu)&=&K^{-1} \times
\sum_{a\neq b} \big[ \tilde W(\u_i - \u_{a}) \, \tilde W^{*}(\u_i -
  \u_{b})\, \nonumber \\
 &\times& \V (\u_{a},\nu) \, \V^{*}(\u_{b},\nu + \Delta\nu) \big]
\label{eq:2venb}
\end{eqnarray}

 where $K=\sum_{a\neq b}\tilde W(\u_i - \u_{a}) \tilde W^{*}(\u_i -
 \u_{b})$ is a normalization constant and $\u_a$, $\u_b$ refer to the
 different baselines in the observational data. The noise bias is
 avoided by dropping the self-correlations ({\it i.e..} the terms with
 $a=b$). We finally determine $C_{\ell}(\Delta\nu)$ using \citep{ali}

\be 
V_{2}(U,\Delta \nu)=\frac{\pi\theta_w^{2}}{2} \left(\frac{\del \I}{\del
  T}\right)^2 C_{\ell}(\Delta \nu) \,Q(\Delta \nu) \,.
\label{eq:cell}
\e 

where $Q(\Delta \nu)$ is a slowly varying function of $\Delta\nu$
which accounts for the fact that we have treated $\theta_w^{2} $ and
$(\del \I/\del T) $ as constants in our analysis.  We have used
$Q(\Delta \nu)=1$ which introduces an extra $\Delta \nu$ dependence in
the estimated $C_{\ell}(\Delta \nu)$ .  This, we assume, will be a
small effect and can be accounted for during foreground removal. The
gridded data has been binned assuming that the the signal is
statistically isotropic in $\u$. We finally invert equation
\ref{eq:cell} to determine the angular power spectrum

\begin{equation}
C_{\ell}(\Delta \nu)=6.8 \times 10^{8} \times \left(\frac{1\,
  rad}{\theta_w}\right)^2 \times \left(\frac{1\,{\rm  MHz}}{\nu}\right)^4
\times \left(\frac {|V_2(\u,\Delta \nu)|}{{\rm Jy}^2}\right)
\label{eq:v2b}
\end{equation}

where, $\theta_w$ is in radians, $\nu$ is the central observing
frequency in MHz and the $C_{\ell}(\Delta \nu)$ is in ${\rm mK}^2$.

The measured $V_2({\rm U},\Delta \nu)$ will, in general, have a real
and imaginary parts (eg. Figure \ref{fig:v2}).  The expectation value
of $V_2({\rm U},\Delta \nu)$ (eq. \ref{eq:cell}) is predicted to be
real, and the imaginary part is predicted to be zero . We use the real
part of the measured $V_2({\rm U},\Delta \nu)$ to estimate
$C_{\ell}(\Delta \nu)$ using eq.  (\ref{eq:v2b}). A small imaginary
part, however, arises due to the noise in the individual
visibilities. The fact that we are correlating two slightly different
visibilities also makes a small contribution to the imaginary part.
Figure \ref{fig:v2} shows the real and imaginary parts of the measured
$V_2({\rm U},\Delta \nu=0)$ as a function of ${\rm U}$ for all the four
fields that we have analyzed. As expected, the imaginary part is much
smaller than the real part.  We use the requirement that the imaginary
part of $V_2({\rm U},\Delta \nu)$ should be small compared to the real
part as a consistency check of our analysis. It also establishes that
we are truly measuring a sky signal, and our results are not dominated
by noise or other local effects in the individual visibilities.

\begin{figure}
\includegraphics[width=80mm,angle=-90]{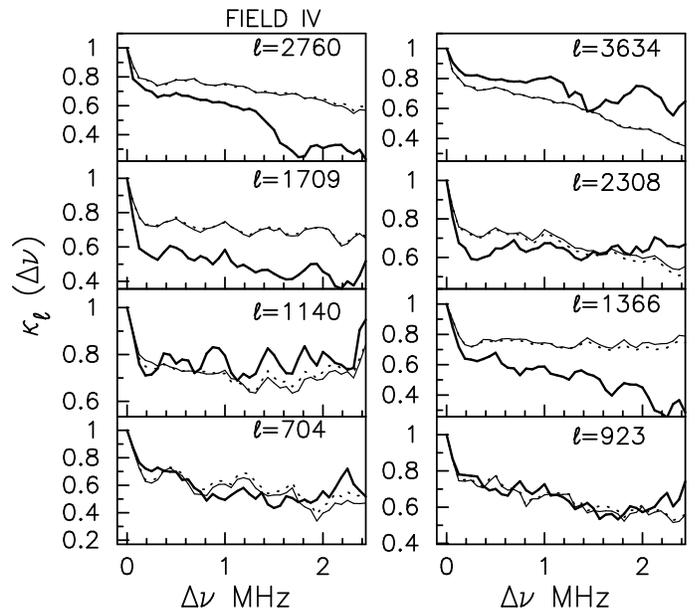}
\caption{The measured $\kappa_{\ell}(\Delta \nu)$ as a function of
  $\Delta \nu$ for the different $\ell$ values as shown in each panel,
  for FIELD IV. The thick solid, thin solid and dotted curves show
  results for no tapering, and tapering with ${\rm f}=0.8 $ and $0.6$
  respectively. }
\label{kappaF4}
\end{figure}

We next investigate whether tapering the sky response of the PB
actually mitigates the oscillations reported in earlier work at $150
\, {\rm MHz}$ (\citealt{ali}, FIELD IV of this paper).  For this
purpose, we introduce the dimensionless decorrelation function
$\kappa_{\ell}(\Delta\nu)=C_{\ell}(\Delta \nu)/C_{\ell}(0)$ which has
the maximum value $\kappa_{\ell}(\Delta \nu)=1$ at $\Delta \nu=0$, and
is in the range $\mid \kappa_{\ell}(\Delta \nu) \mid \le 1$ for other
values of $\Delta \nu$.  We use $\kappa_{\ell}(\Delta \nu)$ to
quantify the $\Delta \nu$ dependence of $C_{\ell}(\Delta \nu)$ at a
fixed value of $\ell$.  Figure \ref{kappaF4} shows
$\kappa_{\ell}(\Delta \nu)$ for FIELD IV with and without the
tapering.  We have considered ${\rm f}=0.6$ and $0.8$ which
respectively correspond to a tapered sky response with FWHM
$2.3^{\circ}$ and $3.04^{\circ}$ as compared to the untapered PB which
has a FWHM of $3.8^{\circ}$ at $150 \, {\rm MHz}$.  The oscillatory
pattern is distinctly visible when the tapering is not applied.  For
most values of $\ell$, the oscillations are considerably reduced and
are nearly absent when tapering is applied. We do not, however, notice
any particular qualitative trend with varying ${\rm f}$. The tapering,
which has been implemented through a convolution, is expected to be
most effective in a situation where the $uv$ plane is densely sampled
by the baseline distribution. Our results are limited by the patchy
$uv$ coverage of the observational data that is being analyzed here.
This also possibly explains why some small oscillations persist even
after tapering is applied.

The results for the other fields are very similar to FIELD IV, and we
have not shown these here. We have used a tapered PB with $f=0.8$ for
all the fields in the entire subsequent analysis.

\section{The Measured MAPS $C_{\ell}(\Delta \nu)$}
\label{maps}

\begin{figure*}
\psfrag{A}[c][c][1][0]{$\ell$}
\psfrag{B}[c][c][1][0]{$\Delta\nu \, \, {\rm MHz}$}
\includegraphics[width=.32\textwidth, angle=-90]{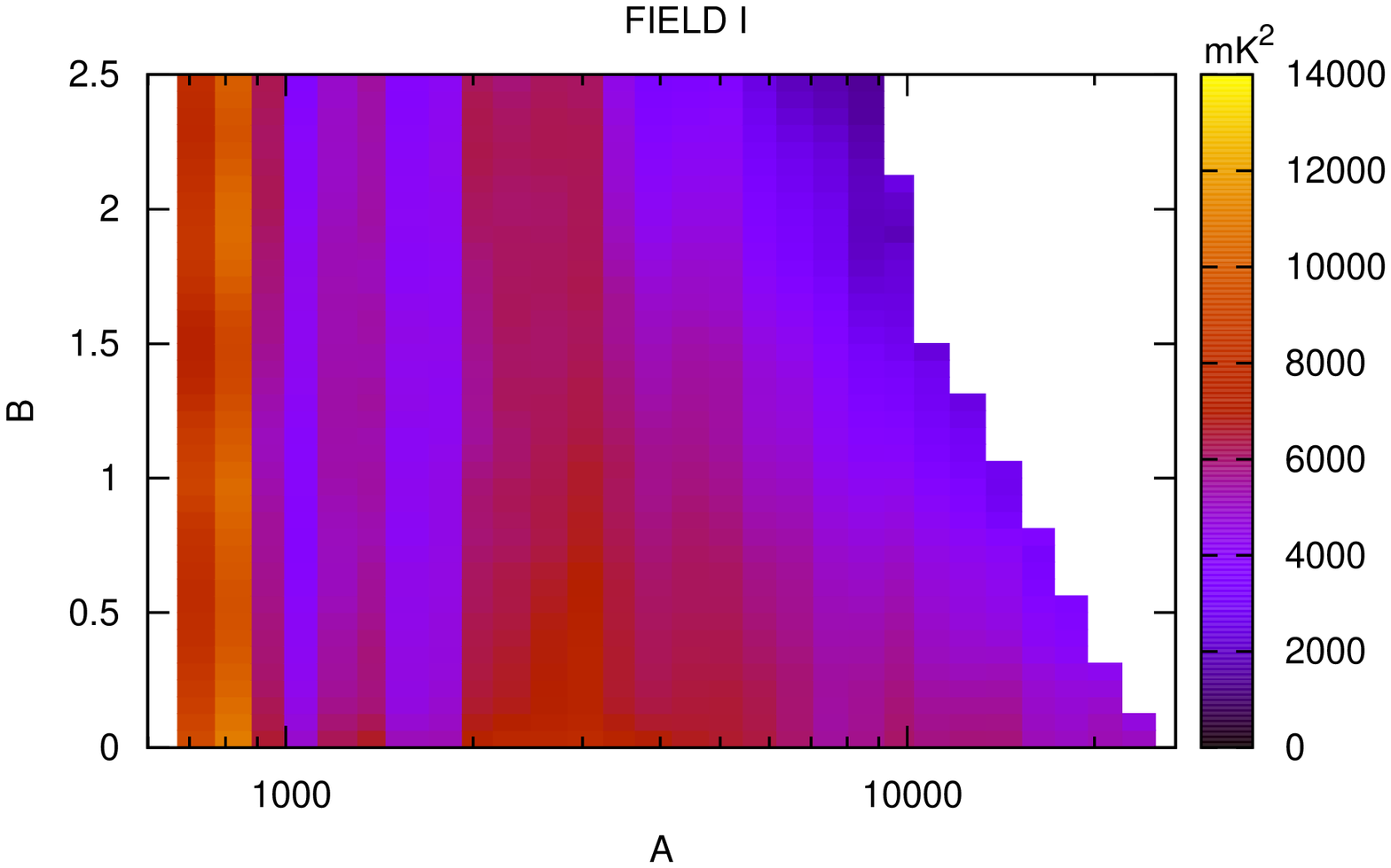}
\includegraphics[width=.32\textwidth, angle=-90]{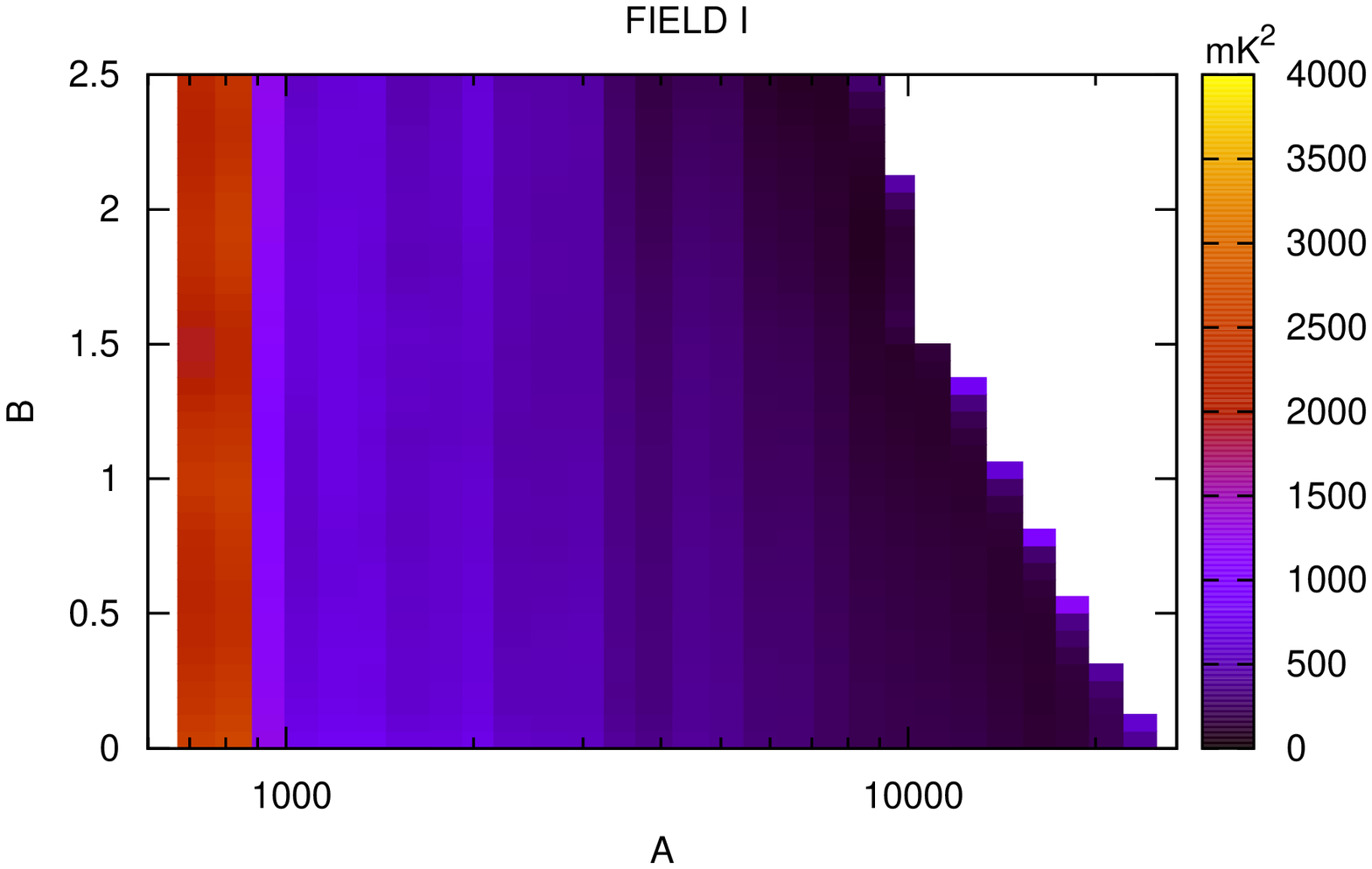}
\includegraphics[width=.32\textwidth, angle=-90]{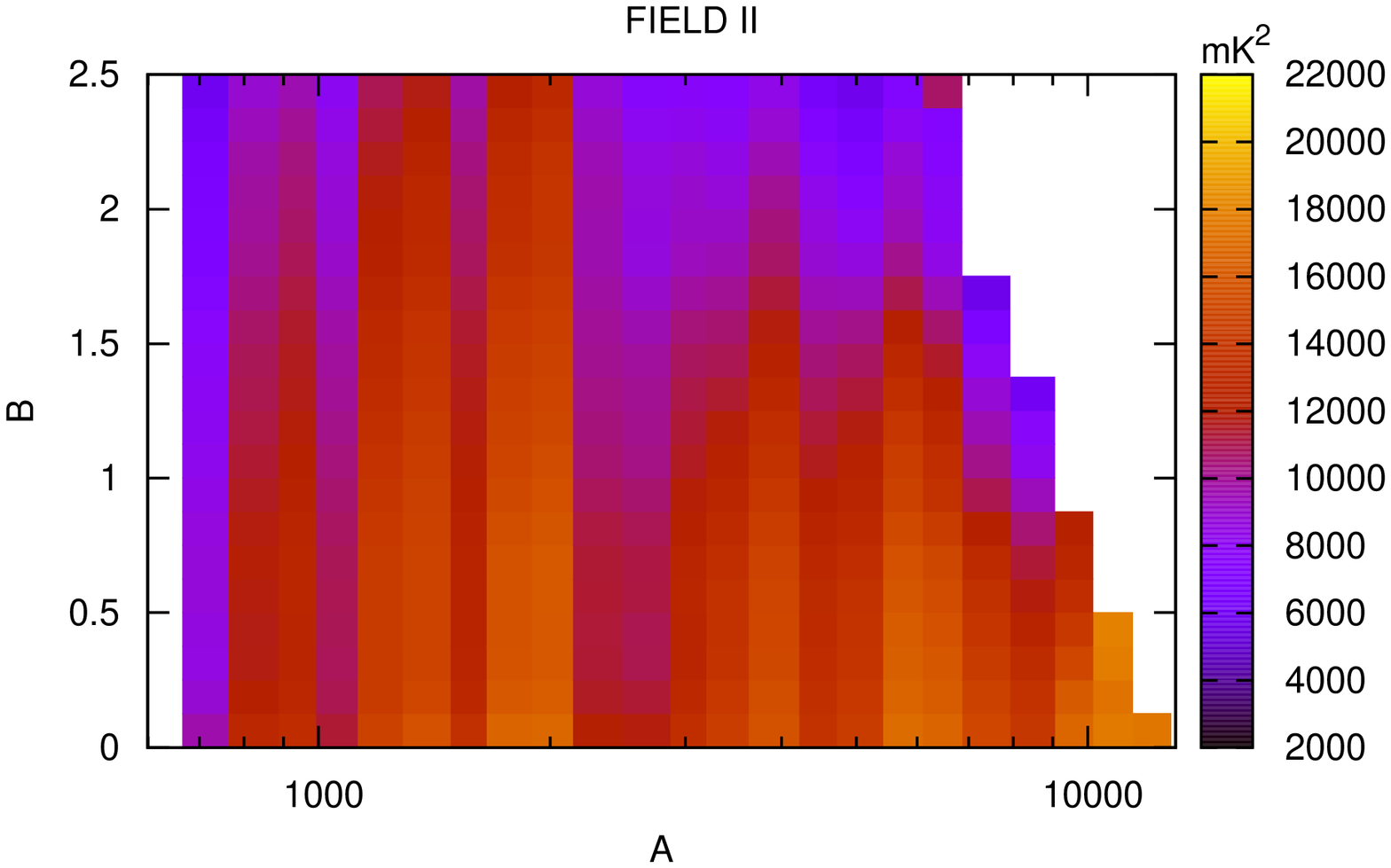}
\includegraphics[width=.32\textwidth, angle=-90]{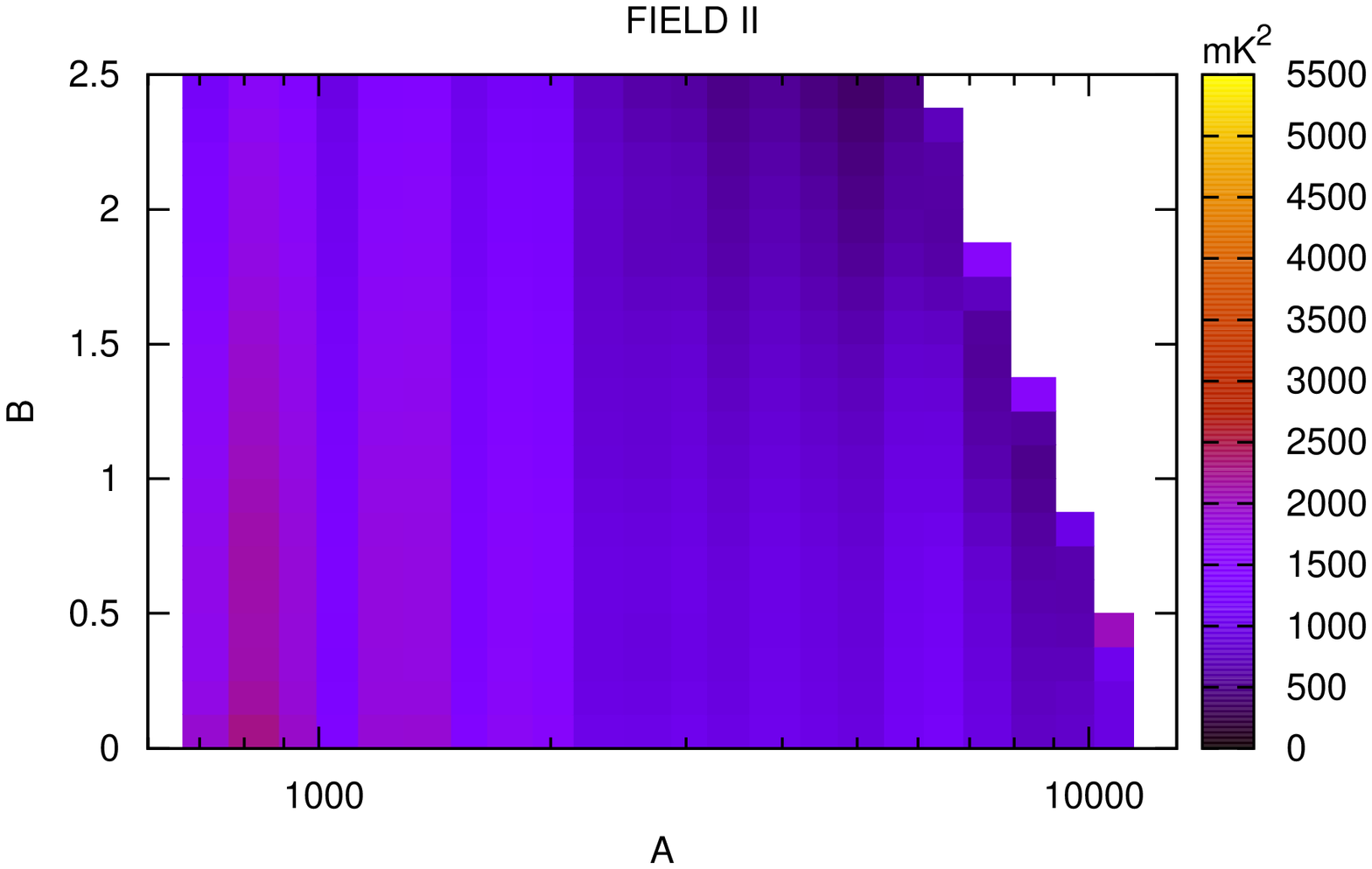}
\caption{The left and right panels show the measured $C_{\ell}(\Delta
  \nu)$ and the $1-\sigma$ error-bars respectively. The white regions
  have been blanked as  no data points are available.} 
\label{fig:clcont}
\end{figure*}

Figure \ref{fig:clcont} shows the measured $C_{\ell}(\Delta \nu)$ and
$1-\sigma$ error estimates for two of our observed fields that we have
analyzed.  We note that the $1-\sigma$ error estimate has two
different contributions, the cosmic variance and system noise, added
in quadrature. In our observations the error is mainly dominated by
the cosmic variance due to the limited number of independent
estimates. The system noise makes a smaller contribution.  We have
determined $C_{\ell}(\Delta \nu)$ in the range $0 \le \Delta \nu \le
2.5 \, {\rm MHz}$ and $700 \le \ell \le 2 \times 10^4$ for FIELDS I
and IV.  FIELDS II and III have been observed for shorter time period
(Table \ref{tab:obs_sum}), and the measured $C_{\ell}(\Delta \nu)$
becomes relatively noisy at the large baselines which are sparsely
sampled in our observational data.  The $\ell$ range has been
restricted to $700 \le \ell \le 10^4$ for FIELDS II and III.  This
corresponds to the angular scales $64^{''}$ to $15^{'}$.

A visual inspection of Figure (\ref{fig:clcont}) shows certain regions
(large $\ell$ and $\Delta \nu$) where we do not have estimates of
$C_{\ell}(\Delta \nu)$, and these regions have been blanked in the
figure. This arises due to a combination of two factors. To start
with, the large baselines are sparsely sampled in the observational
data. Subsequent flagging of the bad channels further depletes the
data. These two factors together lead to a situation where there is
practically no data at large $\Delta \nu$ separations in the large
baselines. It is reassuring to note that this effect is very similar
in all the four FIELDS, of which one was manually flagged and the rest
underwent automated flagging. This establishes that this effect is not
an artifact introduced by the flagging technique.

\begin{figure*}
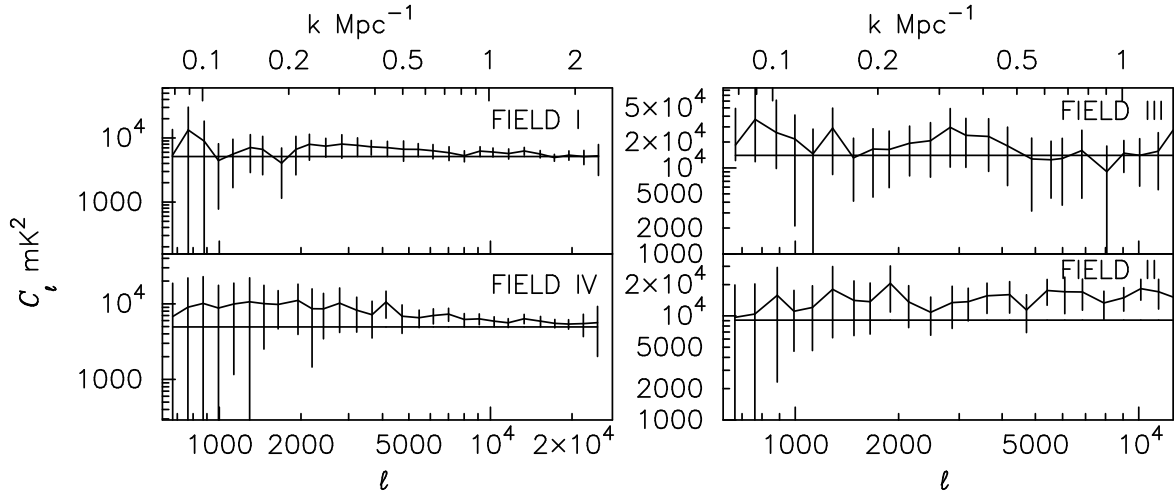

\includegraphics[width=65mm,angle=-90]{cl14.ps}
\includegraphics[width=65mm,angle=-90]{cl23.ps}
\caption{This  shows the measured $C_{\ell}$ as a
  function of $\ell$. The constant straight line shows the   mean
  $C_{\ell}$   corresponding to each field.
The range of Fourier modes $k$ corresponding to
  the $\ell$ modes   probed by our observations are also shown on the
  top $x$ axis of the figure.  The
  $5-\sigma$ error-bars shown here have contributions from both the
  cosmic   variance and system noise.}
\label{fig:cl}
\end{figure*}

\begin{figure*}
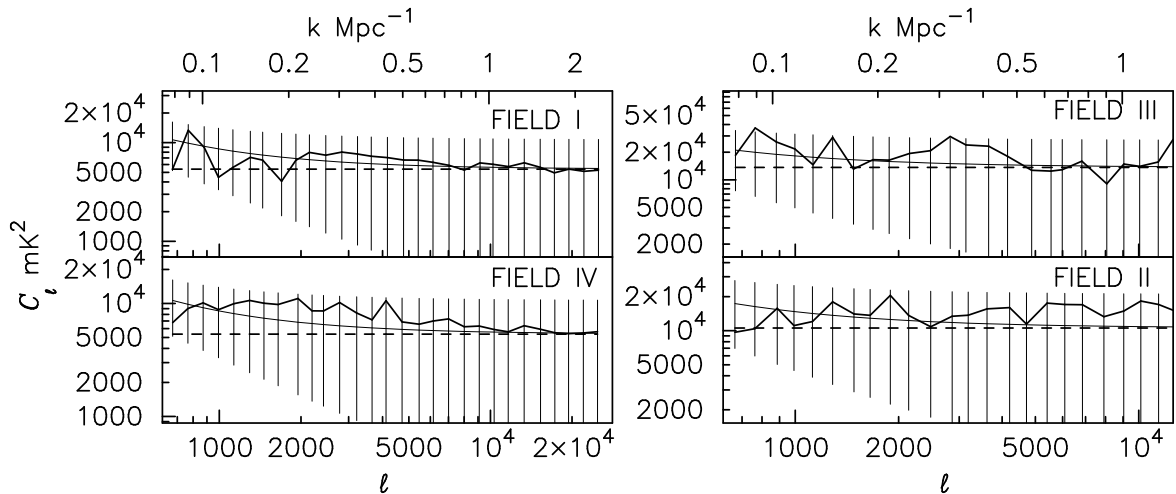

\includegraphics[width=65mm,angle=-90]{cl14_fg.ps}
\includegraphics[width=65mm,angle=-90]{cl23_fg.ps}
\caption{The thick solid line shows the measured $C_{\ell}$ as a
  function of  $\ell$. The thin solid line 
  with $1-\sigma$ error bars shows the total foreground contribution
  based on the model prediction of \citealt{ali}. The dashed line
  shows the contribution  from the   Poisson fluctuation of the
  point   sources which is the most dominant contribution   at large
  $\ell$.   }
\label{fig:clfg}
\end{figure*}

Continuing with the visual inspection of Figure \ref{fig:clcont}, we
do not find any significant pattern in the variation of
$C_{\ell}(\Delta \nu)$ with either $\ell$ or $\Delta \nu$. However,
the values seem to change more along $\ell$ as compared to $\Delta
\nu$.  For each of the four fields, the values of $C_{\ell}(\Delta
\nu)$ are nearly all within $\pm 5 \sigma$ of the mean value of
$C_{\ell}(\Delta \nu)$ which is $5146, 9117, 14,000$ and $4939 \, {\rm
  mK}^2$ for FIELDS I, II, III and IV respectively.  The corresponding
values of $\sigma$ are shown in the right side of each panel of Figure
\ref{fig:clcont}.  The measured $C_{\ell}(\Delta \nu)$ is dominated by
the contribution from extragalactic point sources, and for each field
the mean value of $C_{\ell}(\Delta \nu)$ is determined by the flux
density of the brightest source (${\rm S_c}$) in that field. The
variation in the mean value of $C_{\ell}(\Delta \nu)$ across the
different fields is primarily due to the variation in maximum source
flux density $S_c$ which has values ${\rm S_c}= 0.905 \, {\rm Jy},
1.55 \, {\rm Jy}, 1.9 \, {\rm Jy}, 0.9 \, {\rm Jy} $ in FIELDS I, II,
III and IV respectively (Table \ref{tab:obs_sum}).

We next fix $\Delta \nu$ and study the $\ell$ dependence of $C_{\ell}
\equiv C_{\ell}(\Delta \nu =0)$ (Figure \ref{fig:cl}).  The variation
in the values of $C_{\ell}$ do not seem to exhibit any particular
pattern, and $C_{\ell}$ appears to fluctuate randomly around the mean
values quoted earlier. These fluctuations, as noted earlier, are
nearly all within $\pm 5 \sigma$ of the mean values (Figure
\ref{fig:cl}). We see that the $\ell$ dependence of the measured
$C_{\ell}$ is consistent with random fluctuations arising from the
statistical uncertainties in the measured $C_{\ell}$. Note that the
$5-\sigma$ error bars shown in figure \ref{fig:cl} reflect the
estimated uncertainties arising from both, the cosmic variance and the
system noise.

We model the measured $C_{\ell}$ using the foreground model presented
in \citet{ali}.  The model predictions, shown in Figure
\ref{fig:clfg}, are dominated by the contribution from extragalactic
point sources. At small $\ell$, or large angular scales, the signal is
dominated by the angular clustering of the extragalactic point
sources, whereas the Poisson fluctuation due to the discrete nature of
these sources is the dominant contribution at large $\ell$ which
corresponds to small angular scales. This transition between these two
occurs somewhere in the $\ell$ range $5,000$ to $10^4$, and it is
different in the four fields. The transition shifts to smaller $\ell$
values in the fields which have a larger value of $S_c$. The
uncertainty in the model predictions is dominated by the Poisson
fluctuation of the discrete point sources, and this is nearly constant
across the entire $\ell$ range. The measured $C_{\ell}$ is consistent
with the predictions of the foreground model. Nearly all the measured
$C_{\ell}$ values lie within the $1\sigma$ error bars of the model
predictions.  Other astrophysical sources like the diffuse synchrotron
radiation from our own Galaxy, the free-free emissions from our Galaxy
and external galaxies \citep{shaver} make much smaller contributions,
though each of these is individually larger than the HI signal.

Both simulations \citep{jelic} and analytical predictions
\citep{zald,BA5,cooray, Santos} suggest that at low $\ell$ ($\ell \sim
10^3$) the EoR HI signal is approximately $C^{HI}_{\ell} \sim 10^{-3}
\, - \, 10^{-4} \, {\rm mK^2}$, while at the larger $\ell$ values
($\ell \sim 10^4$) $C^{HI}_{\ell}$ drops to $10^{-5} \, - \, 10^{-6}
\, {\rm mK^2}$.  We find that the measured $C_{\ell}$, arising from
foregrounds, has values around $10^4 \, {\rm mK^2}$ at $\ell \sim
1000$, and it drops by $50 \%$ at the smallest angular scales probed
by our observations $(\ell \sim 10^4)$.  This suggests that the
expected HI signal $C^{HI}_{\ell}$ is ($\sim 10^7$) times smaller than
the measured $C_{\ell}$ arising from foregrounds.

We next shift our attention to the $\Delta\nu$ dependence of the
measured $C_{\ell}(\Delta \nu)$, holding $\ell$ fixed. We have
restricted our analysis to large angular scales ($\ell < 2,000$) where
the HI signal is predicted to be relatively stronger, and we have a
higher chance of an initial detection.  We find that for nearly all
the $\ell$ values for FIELD I and IV the variation in $C_{\ell}(\Delta
\nu)$ with $\Delta \nu$ is roughly between $4\times10^3 \ {\rm mK}^2$
to $10^4 \ {\rm mK}^2$, whereas for FIELD II and III $C_{\ell}(\Delta
\nu)$ roughly varies between $6 \times 10^3 \ {\rm mK}^2$ and
$3.5\times10^4 \ {\rm mK}^2$ across the $2.5 \ {\rm MHz}$ $\Delta \nu$
range that we have analyzed. The fractional variation in
$C_{\ell}(\Delta \nu)$ ranges from $\sim$ 20 to 40 percent for all the
fields. Our foreground model prediction shows that the measured $150
\, {\rm MHz}$ sky signal is dominated by extragalactic point
sources. These are believed to have slowly varying, power law
$\nu^{\alpha}$ frequency spectra. We expect the resultant
$C_{\ell}(\Delta \nu)$ to have a smooth $\Delta \nu$ dependence, and
essentially show very little variation over the small $\Delta \nu$
range ($2.5 \, {\rm MHz}$) that we have considered here ($\Delta
\nu/\nu \ll 1$).  For a fixed $\ell$, the cosmic variance is expected
to introduce the same error (independent of $\Delta \nu$) across the
entire band. As a consequence we do not show the cosmic variance for
the $\Delta \nu$ dependence of the measured $C_{\ell}(\Delta \nu)$.

\begin{figure}
\includegraphics[width=85mm,angle=-90]{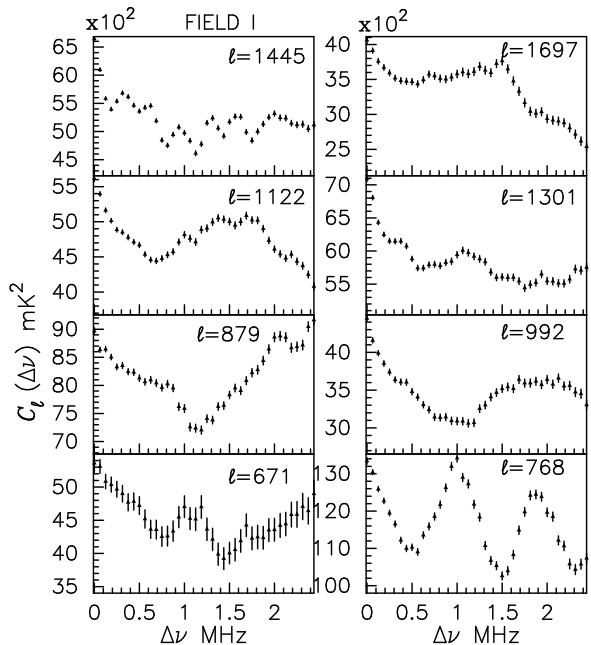}     
\caption{For FIELD I, this  shows the measured $C_{\ell}(\Delta \nu)$ as a
  function of $\Delta \nu$ for the fixed values of $\ell$ shown in the
  panels. The error-bars show $10  \sigma$ system noise.}
\label{fig:clnuf1}
\end{figure}

The measured $C_{\ell}(\Delta \nu)$ is, by and large, found to
decrease with increasing $\Delta \nu$ if we hold $\ell$ fixed.
However, contrary to our expectations, the $\Delta \nu$ dependence is
not completely smooth. There are a few $\ell$ values, particularly in
FIELDS II and III, where the $\Delta \nu$ dependence appears to be
rather smooth. We however have abrupt variations and oscillations in
the $\Delta \nu$ dependence for most of the remaining data.  FIELDS I
and IV, which have a channel width of $62.5 \ {\rm KHz}$, have a
relatively high frequency resolution as compared to FIELDS II and III
which have a channel width of $125 \ {\rm KHz}$.  For FIELDS I and IV,
an oscillatory pattern is clearly visible in $C_{\ell}(\Delta \nu)$ at
nearly all the $\ell$ values.  These oscillations, however, are not
noticeable in FIELDS II and III which have a lower frequency
resolution.  It is possible that the oscillations are actually also
present in FIELDS II and III, but we are missing them due to the lower
frequency resolution in these two fields. To test this we have also
analyzed FIELDS I and IV at a lower frequency resolution by collapsing
the original data. We find that the oscillations seen in FIELDS I and
IV are somewhat reduced, but they can still be made out at the lower
frequency resolution of $125 \ {\rm KHz}$. We find that though the
oscillations in all the fields have reduced considerably after
tapering the sky response (Section \ref{sec3}), a residual oscillatory
patterns still persist.  The oscillations are most pronounced at the
lowest $\ell$ values where the period of oscillation varies between
$\sim 1\, {\rm MHz}\, - \, 2 \, {\rm MHz}$ (Figures \ref{fig:clnuf1}).
The period and amplitude of these oscillations both decrease with
increasing $\ell$.  The exact cause of this residual oscillatory
pattern is, at the moment, not known. We recollect that the sky
tapering is implemented through a convolution whose efficiency depends
on the $uv$ coverage of the baselines. It is possible that the
residual oscillations are a consequence of the finite and sparse $uv$
coverage of our data.

The oscillations and abrupt changes in the $\Delta \nu$ dependence of
$C_{\ell}(\Delta \nu)$ posse a severe impediment for foreground
removal. Foreground removal relies on the key assumption that the
foregrounds have a $\Delta \nu$ dependence which is distinctly
different from that of the HI signal. The foreground contributions are
expected to vary slowly with increasing $\Delta \nu$, whereas the HI
signal is expected to decorrelate rapidly well within $\Delta\nu \le
0.5 \ {\rm MHz}$ \citep{BS1,BA5,ghosh1}. It is then possible to remove
the foregrounds by modelling and subtracting out any component that
varies slowly with increasing $\Delta \nu$. For example, it is
possible to use low order polynomials to model the $\Delta \nu$
dependence of the measured $C_{\ell}(\Delta \nu)$ at fixed valued of
$\ell$, and use these to subtract out the foreground contribution
\citep{ghosh1,ghosh2}. It is quite evident that the oscillations and
abrupt changes in the $\Delta \nu$ dependence of the measured
$C_{\ell}(\Delta \nu)$ posse a severe challenge for this technique, or
any other technique that is based on the assumption that the
foregrounds vary smoothly with frequency. It is relevant to note that
\citet{jelic,harker} have shown that the foreground subtraction using
polynomial fitting can easily cause over-fitting, in which we fit away
some of the cosmological signal, or under-fitting, in which the
residuals of the foreground emission can overwhelm the cosmological
signal. Therefore, one needs to use non-parametric methods that allow
data to determinate their shape without selecting any a priori
functional form of the foregrounds \citep{harker,chapman}.

We would like to point out that a possible line of approach in
foreground removal is to represent the sky signal as an image cube
where in addition to the two angular coordinates on the sky we have
the frequency as the third dimension
\citep{jelic,harker,chapman,zaroubi}. For each angular position,
polynomial fitting is used to subtract out the component of the sky
signal that varies slowly with frequency. The residual sky signal is
expected to contain only the HI signal and noise
\citep{jelic,bowman09,Liu1,harker,harker10}. \citet{Liu1} showed that
this method of foreground removal has problems which could be
particularly severe at large baselines if the $uv$ sampling is
sparse. We would also like to point out that in a typical GMRT
observation the $uv$ plane is not completely sampled and due to this
limited baseline coverage there exist correlated noise in the image
plane which is a problem for estimating the power spectrum in the
image plane (for details \citet{prasunth}, Appendix A). Although, the
upcoming low frequency radio telescopes such as the LOFAR, MWA
etc. will have a much dense $uv$ coverage compared to GMRT, where in
the core region most of the baselines relevant for the EoR experiments
will be sampled and the foreground removal technique in
image-frequency space can also be applied
\citep{jelic,harker,chapman,zaroubi}.

\begin{figure}
\includegraphics[width=85mm,angle=-90]{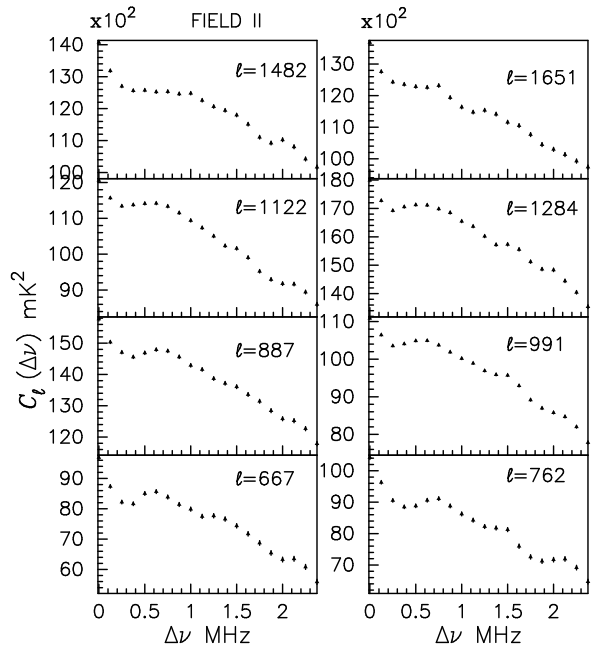}     
\caption{For FIELD II, this  shows the measured $C_{\ell}(\Delta \nu)$ as a
  function of $\Delta \nu$ for the fixed values of $\ell$ shown in the
  panels. The error-bars show $10  \sigma$ system noise.}
\label{fig:clnuf2}
\end{figure}

\section{Point sources}
\label{ps}

The discrete sources seen in Figure \ref{fig:fields} dominate the $150
\, {\rm MHz}$ radio sky at the angular scales probed in our
observations. These sources are typically smaller than the angular
resolution of our observations. It is now well accepted that these are
extragalactic, being mainly associated with active galactic nuclei
(AGN). It is of considerable interest to study the properties of these
sources \citep*{dunlop90,jackson99,peacock,Seymour,Garn}.

There currently exist several radio surveys at $150 \,{\rm
  MHz}$ like the 3CR survey \citep{Bennett}, the 3CRR catalogue
\citep{laing}, the 6C \citep*{hales,waldram} and the 7C survey
\citep{hales07}, which cover large regions of the sky.  The 6C and 7C
surveys have angular resolutions of $\sim 4.2^{'}$ and $\sim 70^{''}$
respectively, and the limiting source flux density of these surveys is
$\sim 100\, {\rm mJy}$.

Point sources are the most dominant foreground component and it
creates a major problem for detecting the 21-cm signal from the
redshifted HI emission in the $\ell$ range $10^3 - 10^4$ probed here.
\citet{dmat1} have used the results of the previous surveys to
estimate the point source contribution to the foregrounds at $150 \,
{\rm MHz}$.  However, these have a limiting source flux density of
$\sim 100\, {\rm mJy}$, and the extrapolation to fainter sources is
rather uncertain.  The GMRT has an angular resolution of $\sim 20''$
at $150 \, {\rm MHz}$. We are able to achieve a rms noise of around
$1.3\, {\rm mJy/Beam}$ in the observations reported here. The GMRT is
currently the only instrument capable of achieving this level of
sensitivity in terms of angular resolution and rms noise. We note that
there are relatively few GMRT results regarding the radio source
population at $150 \, {\rm MHz}$ \citep*{ishwara,intema} with
sensitivity comparable to that reached in our observation. Recent
simulations \citep*{bowman09,liu09b} also indicate that point sources
should be subtracted down to a $\sim 10-100\, {\rm mJy}$ threshold in
order to detect the EoR signal. In this section we use our $150 \,
{\rm MHz}$ observations to explore the population of radio sources in
our observed fields down to the detection limit of $\sim 9\, {\rm
  mJy}$.

In order to detect the HI signal, it is very important to correctly
identify the point sources and subtract these out at a high level of
precision \citep*{Bernardi11,pindor11}. It is quite evident that a
template of detected point sources above a given threshold level is an
essential part of foreground removal where we have noticed that the
Poisson and clustering component of the point sources are the most
dominating foreground component at our angular scales of analysis. We
first focus on the brightest source in each of the four fields that we
have analyzed. In each field, the brightest source (Table
\ref{tab:img_sum}) alone contributes around $10 \%$ of the total
measured $C_{\ell}$. We consider the brightest source as a test case
to investigate how well it is possible to image and subtract out the
point sources.

\begin{figure*}
\includegraphics[width=75mm]{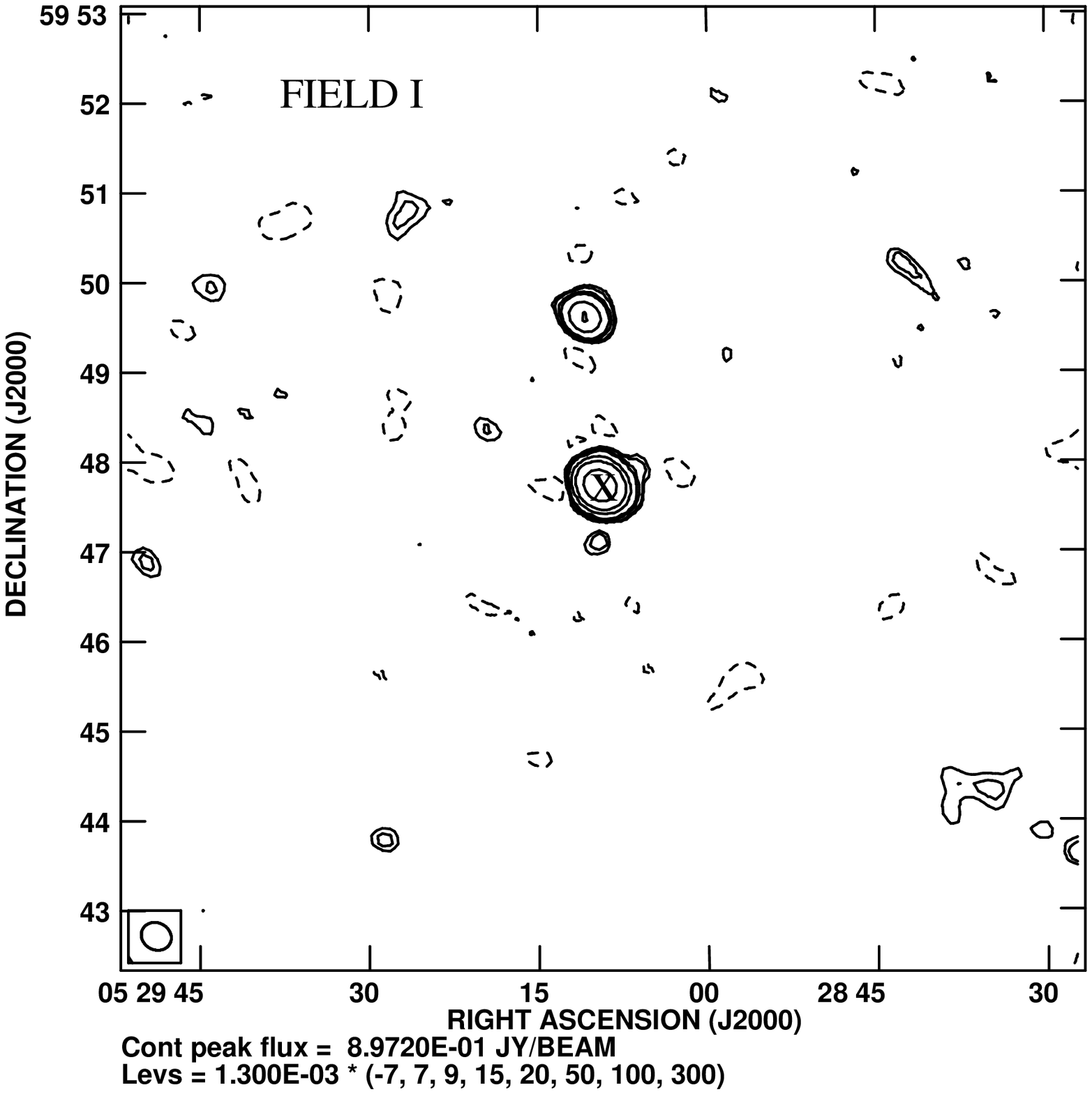}
\includegraphics[width=75mm]{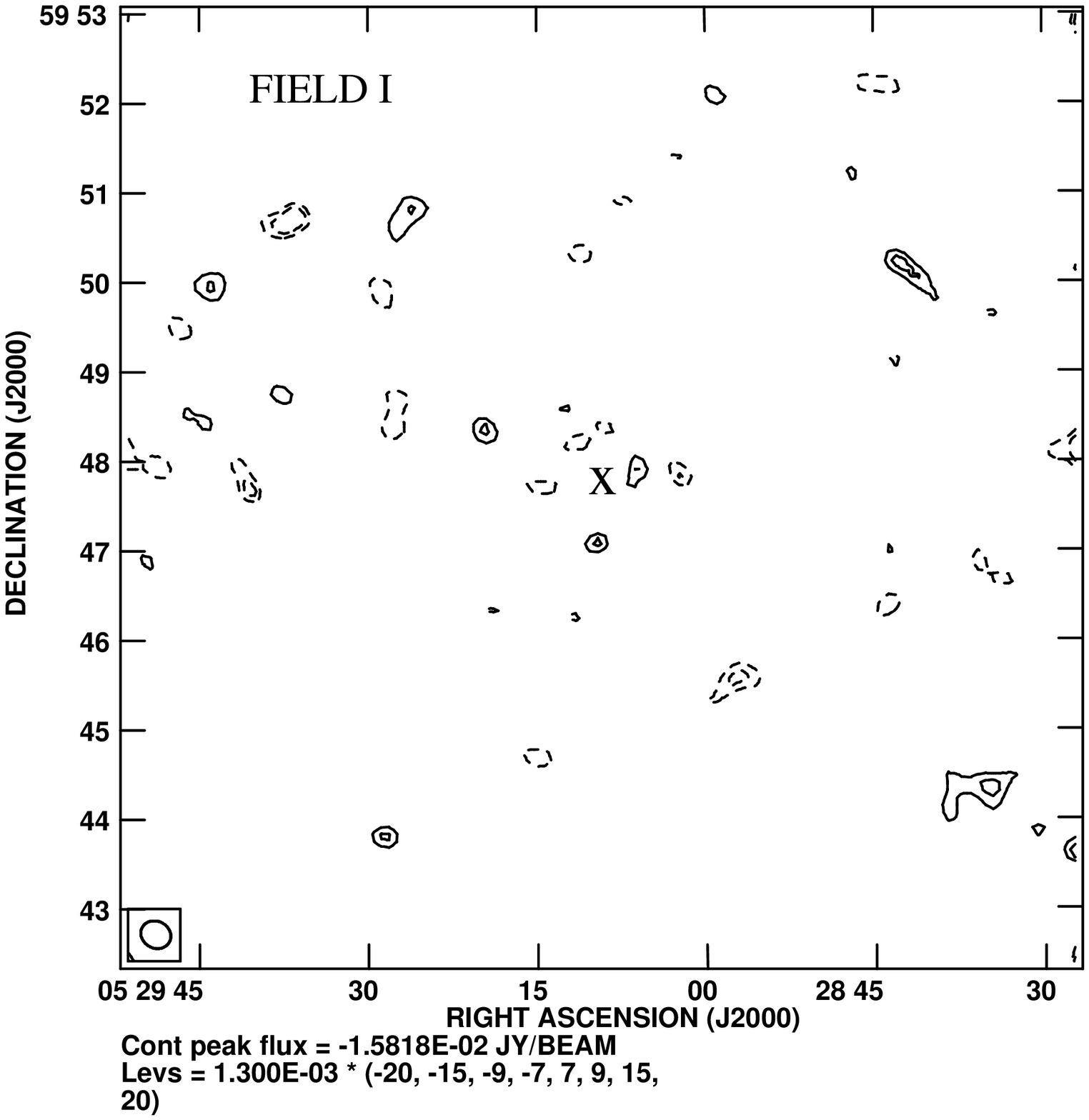}
\includegraphics[width=75mm]{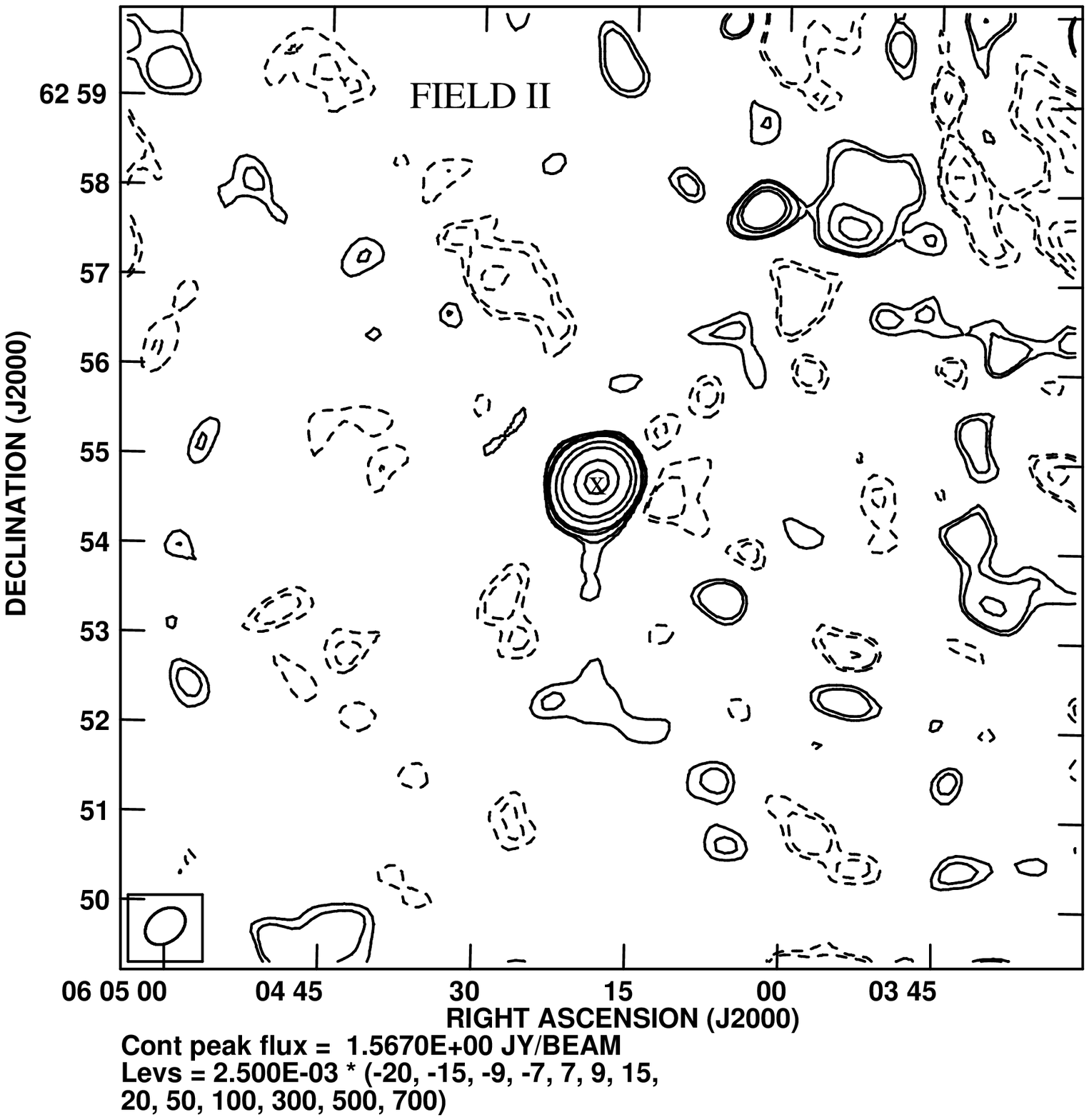}
\includegraphics[width=75mm]{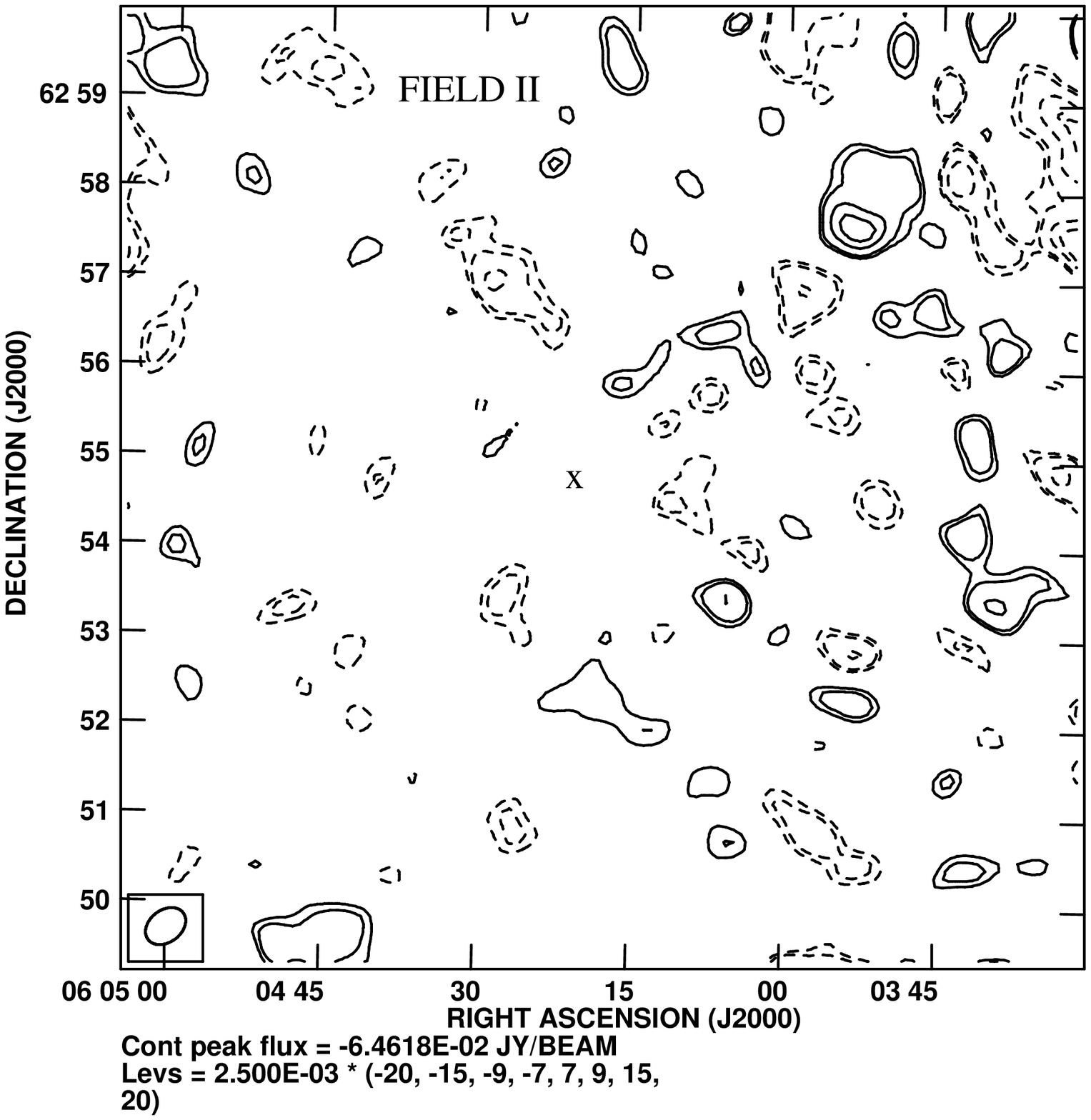}
\caption{ This figure shows the position around the brightest source
  (marked with `X') for FIELD I and II before (left panel) and after
  (right panel) the source is subtracted.}
\label{fig:bs1}
\end{figure*}

We note that FIELDS I and IV have a relatively longer on-source
observation time in comparison to FIELD II and III, and this is
reflected in the fact that FIELDS I and IV have a lower noise level
and higher sensitivity in comparison to FIELD II and III (Table
\ref{tab:img_sum}).  The left panel of Figure \ref{fig:bs1} shows a
more detailed view of the brightest source for the most sensitive and
the least sensitive fields among the four fields that we have imaged.
For making these images, we have applied the appropriate phase shifts
so as to bring the brightest source to the phase center of the image.
The ratio of the peak flux density to the rms noise (Peak/Noise, Table
\ref{tab:img_sum} ) has a maximum value of $700$ in FIELD I, and a
minimum value of ($422$) in FIELD III.  The brightest source in each
of our fields is also found to be accompanied by several regions of
negative flux density. These are presumably the result of residual
phase errors which were not corrected in our self calibration process.
We find that the Peak negative flux density is $14 \, {\rm mJy/Beam}$
in FIELD I, and $47 \, {\rm mJy/Beam}$ in FIELD IV.  These correspond
to $1.5 \%$ and $5.3 \%$ of the Peak flux density in the respective
FIELDS, the values lie within this range for the two other fields
(Table \ref{tab:img_sum}).  In addition to the pixels with negative
flux densities, we also see several regions of positive flux densities
well above the $5\sigma$ noise levels. Both the positive and negative
regions are imaging artifacts which are possibly the outcome of
calibration errors.  The self-calibration steps implemented in the
earlier stage of the analysis have considerably reduced the imaging
artifacts. The artifacts, however, not entirely removed through self
calibration.  A visual inspection of the image of the brightest source
in FIELD I (top left panel of Figure \ref{fig:bs1}) shows that there
are around $10$ distinct features with flux densities around $7 \,
{\rm mJy}$ arising from imaging artifacts.  The artifacts effectively
increase the local rms noise in the vicinity of the brightest source,
and reduces the dynamic range of the image.

We next consider how well we can subtract out the brightest source in
the four fields that we have analyzed. We model the brightest source
using the Clean Components of the continuum image.  The visibilities
corresponding to these clean Components were subtracted from the
original full frequency resolution $uv$ data using the AIPS task
UVSUB. The right panel of Figure \ref{fig:bs1} shows the continuum
image made with the residual visibilities after the brightest source
was subtracted out. We find that the peak residual has values
$(-15,-64,-54,34) \, {\rm mJy/Beam}$ which correspond to
$(1.6,4.0,3.1,3.8) \%$ of the original source that was subtracted out
from the four fields respectively. FIELD I has the least residuals and
the image is largely featureless in comparison to FIELD II which has
the maximum residuals.

\begin{figure*}
\includegraphics[width=55mm,angle=-90]{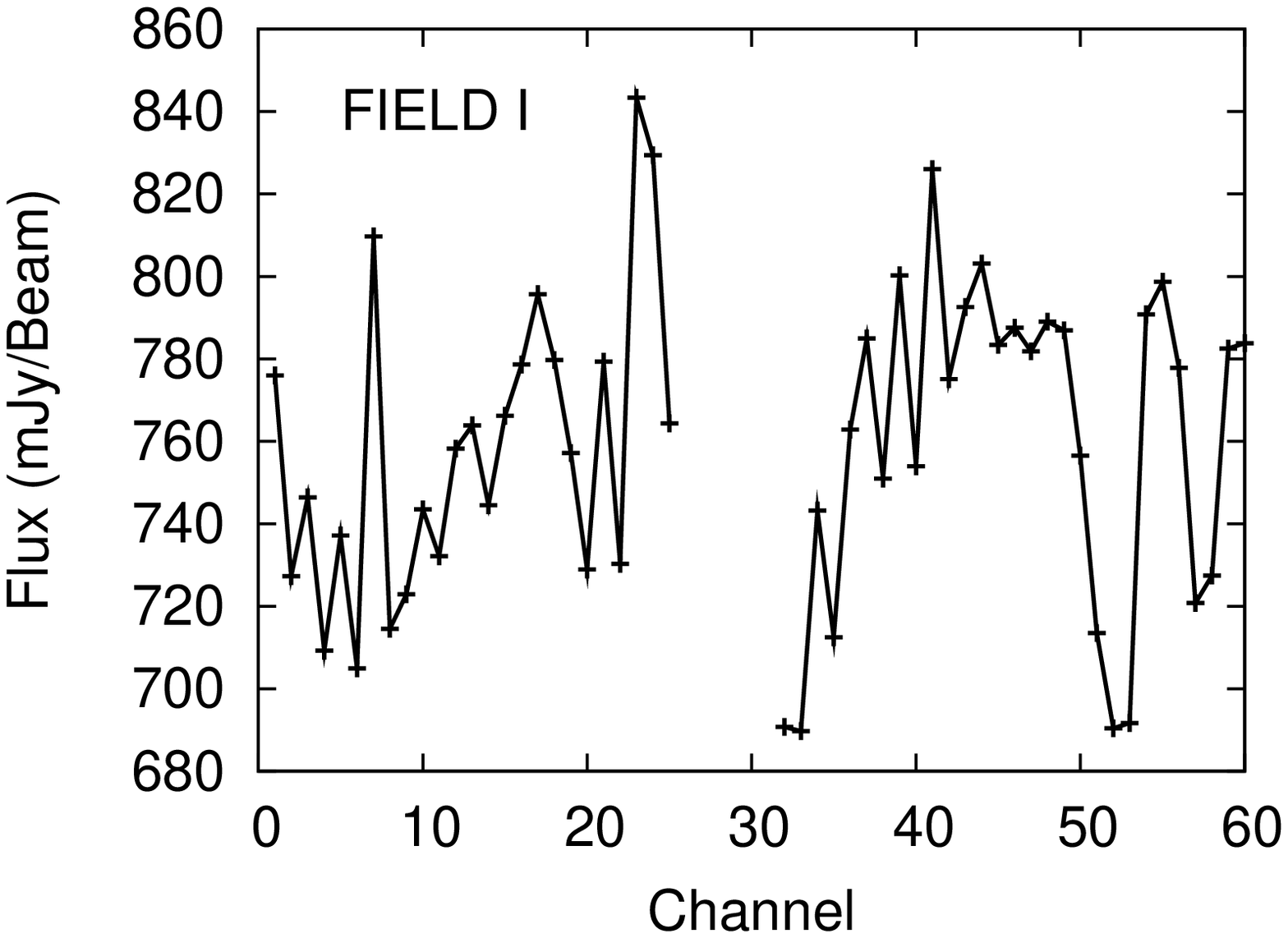}
\includegraphics[width=55mm,angle=-90]{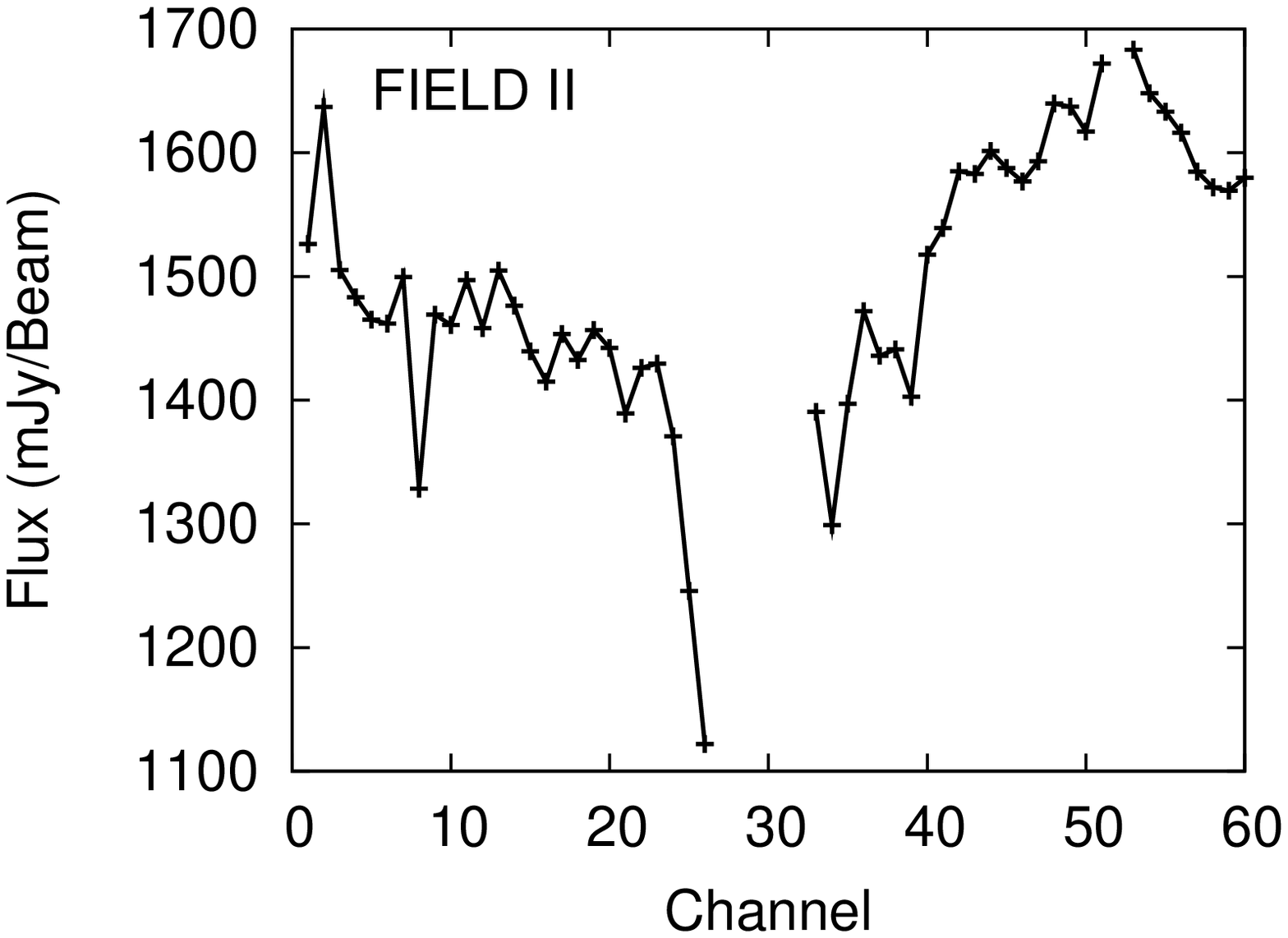}
\caption{ The figure shows the Flux/Beam vs channel (Frequency) plots
  at the brightest source position for two fields. Note that some
  channels were heavily corrupted by broad-band RFIs and we flagged
  those channels which are shown as `gap' in FIELD I and II.}
\label{fig:spec}
\end{figure*}

The continuum image used for source subtraction does not take into
account the possibility that the variation in the source flux across
the observational band.  In principle, we do not expect a significant
flux variation across the relatively small frequency bands in our
observation.  Figure \ref{fig:spec} shows the channel spectra through
the brightest pixel in two of our observed fields. Contrary to our
expectation, we find $15 \, \%$ to $30 \, \%$ variations in the
spectra. Further, these variations are not smooth and they show random
and abrupt variations. The channels where data is missing have been
flagged to avoid broad band RFI.  This spectral variation is mainly
due to errors in the bandpass calibration. It is very clear that it
will not be very effective to model the observed spectral variation of
the source using a smooth polynomial or power law.  It may be a better
strategy to make separate images at each frequency channel and
individually subtract out the CLEAN component from each channel of the
$uv$ data, provided the signal to noise ratio is sufficient for single
channel images.

\begin{figure*}
\includegraphics[width=75mm]{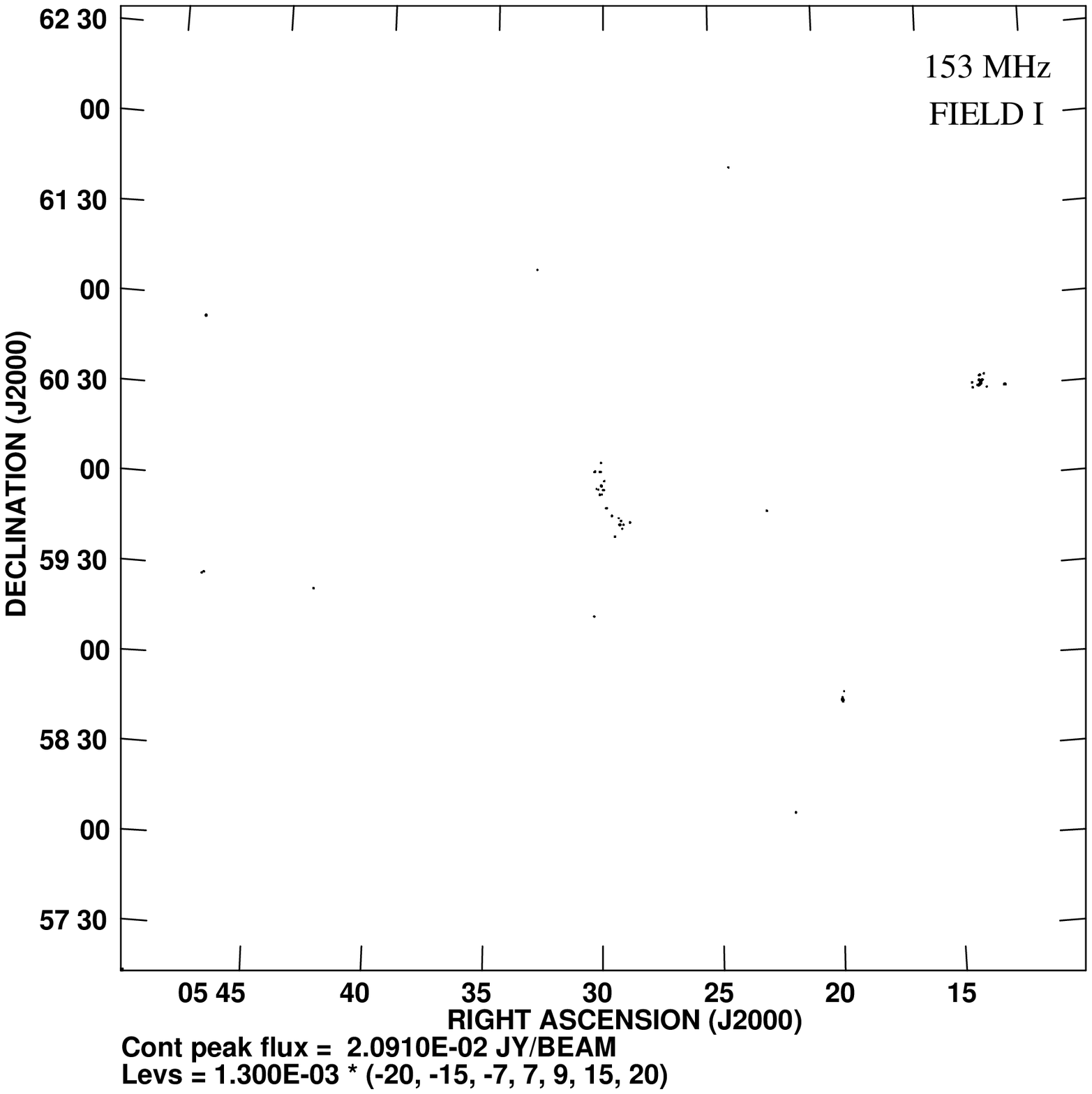}
\includegraphics[width=75mm]{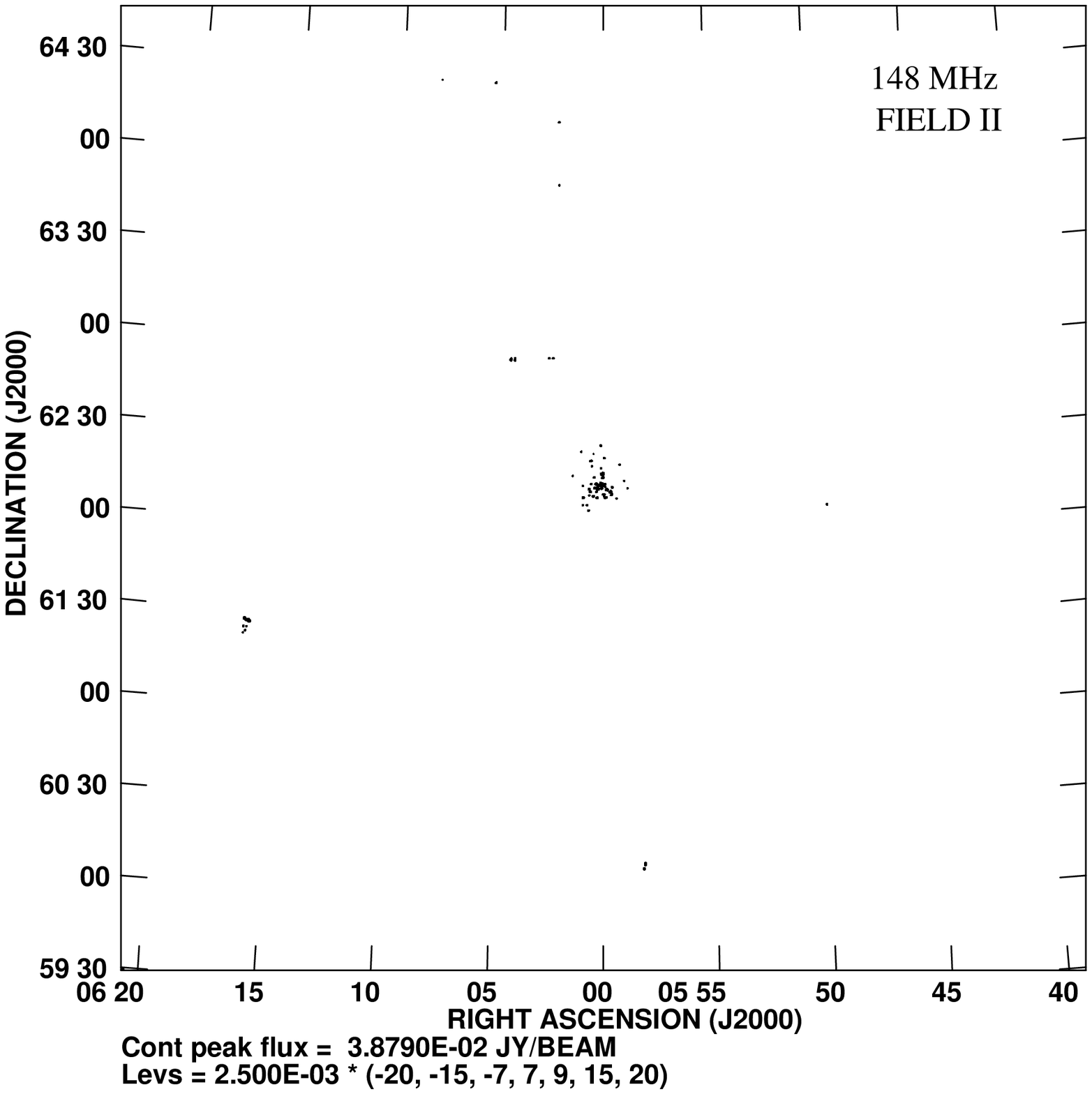}
\includegraphics[width=75mm]{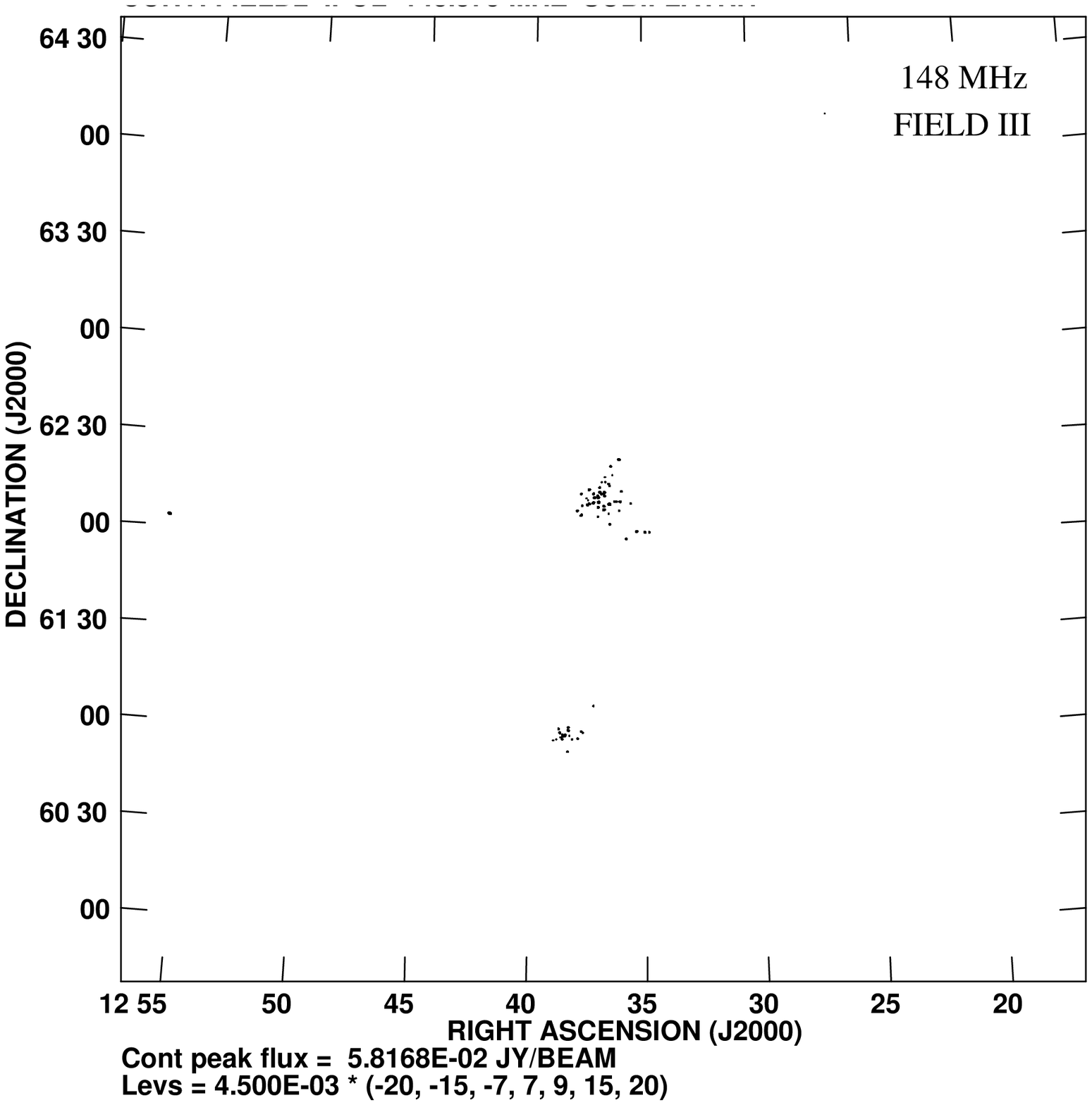}
\includegraphics[width=75mm]{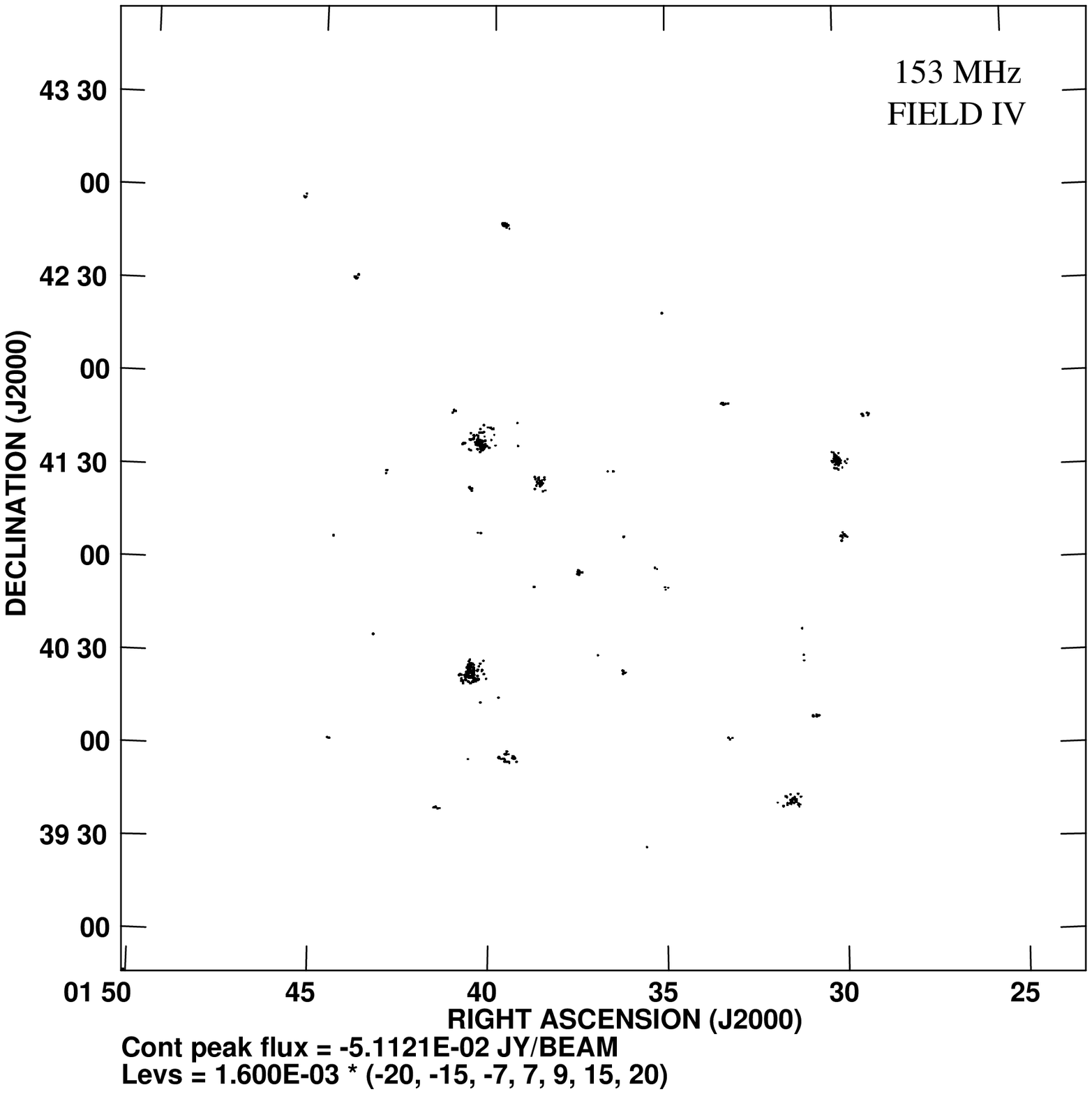}
\caption{This shows the same figure as the Figure \ref{fig:fields}
  except that all the pixels with flux density above $7 \sigma$ noise
  value were visually identified as genuine sources and not artifacts
  have been fitted with clean components and removed from the
  visibility data from which this image was made. It is expected that
  most of the genuine sources have been removed from this data.}
\label{fig:subfields}
\end{figure*}

We next consider source subtraction from the entire field of view.
Pixels with flux density above $7 \sigma$ times the rms noise were
visually identified as sources and subtracted out using exactly the
same procedure as used for the brightest source.  It is expected that
at this stage most of the genuine sources seen in Figure
\ref{fig:fields} have been removed from the $uv$ data. Figure
\ref{fig:subfields} shows the corresponding images made from the
residual visibilities after source subtraction.  The flux density in
these images are in the range of $(21,39,58,30) \, {\rm mJy/Beam}$ to
$(-14,-29,-51,-51)\, {\rm mJy/Beam}$ respectively for the four fields
that we have analyzed.  Most of the residuals, we find, are clustered
in a few regions in the image. These regions have a typical angular
extent of $15'$ and are centered on the locations of the bright
sources in the FoV.  The residuals are essentially imaging artifacts
that were not modelled using Clean Components.  The rest of the image,
leaving aside a few isolated regions, is largely free of artifacts and
devoid of any visible feature. We find that FIELD I has the highest
sensitivity (lowest rms noise) amongst our observed fields. We also
see that source removal is most effective for this field. We see
(Figure \ref{fig:subfields}) that most of the image is free of
residual structures after source subtraction, except for two small
regions where most of the artifacts are localised. {\it It is quite evident
that {\bf FIELD I} is the best field and we use this for the entire
subsequent analysis.}

\subsection{Differential Source Count}
\label{dsc}

The differential source count plays a very important role in
estimating the power spectrum of point sources.  Further, this also
provides invaluable information as to the nature of the radio sources.
Here we evaluate the differential source count using the discrete
sources identified in FIELD I down to a flux limit of $9 \, {\rm
  mJy}$.

We use the AIPS task SAD to identify and extract the sources from the
continuum image of FIELD I (Figure \ref{fig:fields}). SAD identifies
potential sources based on the peak brightness and fits a Gaussian
model to the sources.  Source identification was restricted within a
radius of $1.45^{\circ}$ from the phase centre, where the primary beam
response is within $50$ per cent of its central value.  We have used a
conservative peak brightness detection limit of $7\sigma$ ($> 9 \,
{\rm mJy}$/Beam ) for the initial source selection using SAD. This
minimizes the number of noise spikes that are spuriously detected as
sources.

As mentioned earlier, the local rms noise increases near the bright
sources due to the presence of imaging artifacts.  We have estimated
the local rms in different parts of the residual image by applying the
AIPS task RMSD. The local rms at each pixel was estimated using an
area of approximately $256 \ \times 256 \ {\rm pixels}$ centered on
that pixel.  Sources with a peak brightness in excess of $5 \sigma$
times the local rms noise were finally selected. Further, we have
visually inspected the regions close to the brightest sources and
discarded sources that appeared to be imaging artifacts.  This ensures
that we do not include any bright imaging artifacts as sources.  The
flux density of all the extracted sources were corrected for the GMRT
primary beam using an eighth order polynomial \citep{kantharia}.

\begin{figure}
\includegraphics[width=75mm,angle=-90]{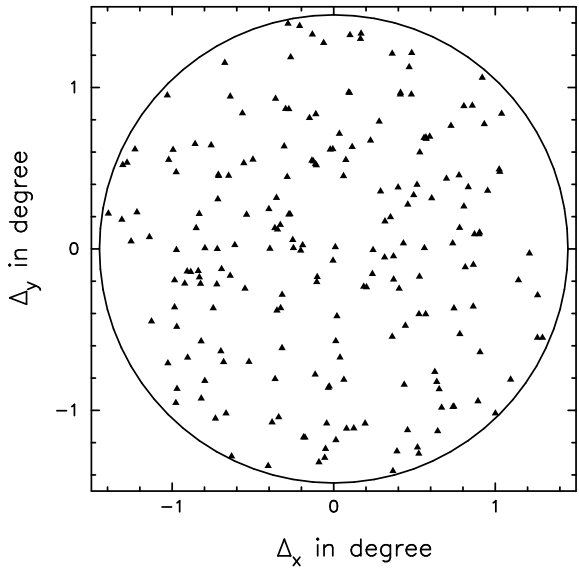}
\caption{The triangles show the 206 sources within a radius of
  $1.45^{\circ}$ from the phase centre. Here, $\Delta_x$ =
  COS(DECC)x(RA-RAC) and $\Delta_y$ = (DEC-DECC) are the angular
  displacements from the phase center RAC and DECC. The RA and DEC
  represents the positions of the different sources.}
\label{radec}
\end{figure}

We have finally identified a total of 206 sources (Figure \ref{radec})
within a 1.45$^{\circ}$ radius from the pointing centre .  The full
source catalogue is given in the online version of the paper (
Appendix \ref{catal}) and is presented in order of increasing RA. For
each source, the catalogue lists the position of the sources in RA,
DEC, peak flux density, local rms noise at the source location, the
integrated flux density and the error in the integrated flux density.

We have used the source catalogue to determine the differential source
count $dN/dS$, the number of sources $dN$ in the flux interval
$dS$. Note that $S$ here refers to the integrated flux density of the
source. To determine $dN/dS$ we have binned the sources in the range
$9.11 \, \le S \le 945.71 \, {\rm mJy}$ into $10$ logarithmic bins in
$S$ (Table \ref{dnds}) and counted the number of sources ($N$) in each
bin.  We next consider two corrections that have to be considered in
interpreting the source counts.  The first is the Eddington bias where
the random noise artificially boosts the number of sources in the
faintest bin.  The $7 \sigma$ detection limit and the subsequent
visual inspection help to avoid spurious sources and minimize the
effect of the Eddington bias.  Further, studies at $610 \ {\rm MHz}$,
\citet{moss} indicate that the Eddington bias only influences the
faintest flux density bin, increasing the source count by
approximately 20$\%$. Since the number of sources in our lowest flux
bin is comparatively small, we decided to make no correction for this
effect.

The second effect is related with the fact that extended sources with
peak brightnesses below the survey limit but integrated flux densities
above this limit would not be detected by our source detection
procedure. It is possible to get an estimate of this effect with the
knowledge of the angular size distribution of the sources as a
function of flux density. This helps to identify the incompleteness
due to extended sources in the estimated source count distribution,
well-known as resolution bias. Recently, \citet{moss} have estimated
that they will miss approximately $3\%$ of the sources due to the
resolution bias at $610 \ {\rm MHz}$. In our case, we do not find a
significant number of extended sources and therefore we choose to make
no correction for this effect.

The actual threshold of flux density for source detection varies
across the image, this being $5$ times the local rms. Consequently,
for each flux bin in Table \ref{dnds} we only have a fraction $f$ of
the image where a source at the faint end of the bin can be
detected. We correct for this by using $N_c=N/f$, the noise corrected
source count, to estimate the differential source count as

\begin{equation}
\frac{dN}{dS} = \frac{N_c}{A\Delta S}\,.
\end{equation}

where $A$ is the total area of the image in steradians, and $\Delta S$
is the width of the flux bin in Jy.  We have used the Poisson error
$\Delta N_c=\sqrt{N_c}$ to estimate the error in the differential
source count.

\begin{figure}
\includegraphics[width=60mm,angle=-90]{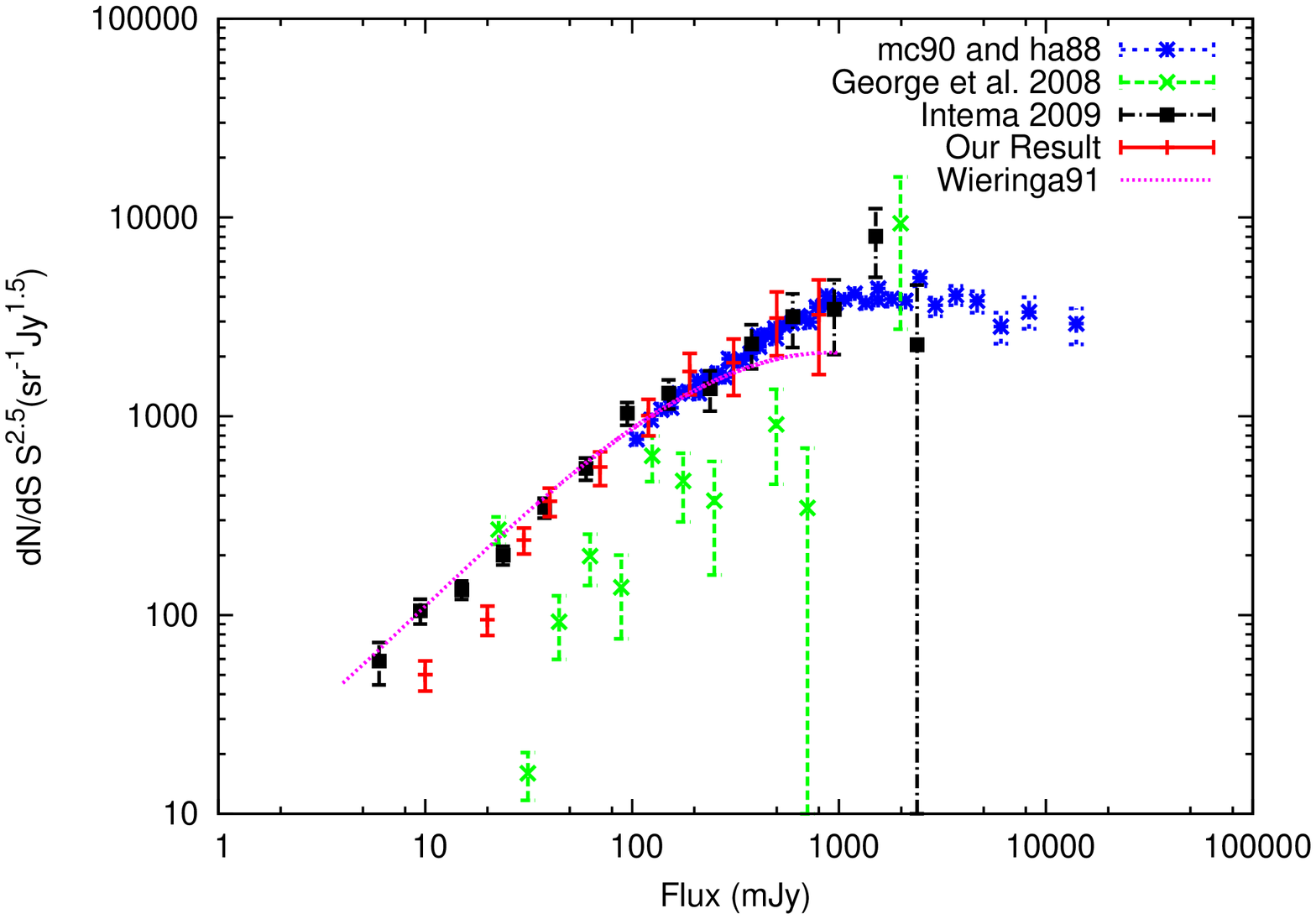}
\caption{The figure shows the differential source counts at $150 \,
  {\rm MHz}$ (mc90 and ha88 (\citealt{McGilchrist} and
  \citealt{hales}), \citet{George}, \citet{intema}, Our Result),
  normalized by the value expected in a static Euclidean universe. A
  steady decrease in the source counts is observed as we approach
  lower flux values. The continuous curve represents the functional
  form of the fit (valid from $4 \, {\rm mJy}$ to $1 \, {\rm Jy}$ ) to
  the source counts by \citet{Wieringa}, scaled down to $150 \, {\rm
    MHz}$ assuming a mean spectral index of -0.70.}
\label{nordnds}
\end{figure}

We have used least square to fit a power law to the differential
source count $dN/dS$ as a function of the mean flux density $S$ of
sources in each bin.  We find the best fit power law to be $dN/dS=
10^{3.75 \pm 0.06} \times S^{-1.6 \pm 0.1}$ for our data. We note that
\citet{dmat1} have found a single power-law fit $dN/dS= 4000 \times
S^{-1.75}$ upto the flux density of $880 \, {\rm mJy}$ which is mostly
consistent within the error bars of our fit. Figure \ref{nordnds}
shows the normalised differential source count $S^{2.5} \times dN/dS$
as a function of the flux density $S$ assuming an Euclidean
Universe. We have compared our results with the $330 \,{\rm MHz}$
source count model from \citet{Wieringa} where the fit to the combined
source count from six different fields (in the range from $4 \, {\rm
  mJy}$ to $1 \, {\rm Jy}$) is given by

\begin{equation}
\log_{10}\left(\frac{dN}{dS}S^{2.5}\right)= 0.976 + 0.6136x + 0.3028x^2
-0.083x^3
\label{eq:war}
\end{equation}

where $x=\log_{10}(S)$ and S is given in mJy.

This source count model was scaled down to our observing frequency of
$150 \, {\rm MHz}$ assuming that the source fluxes can be scaled as $S
\propto \nu^{-\alpha}$. We have tried out various values of the mean
spectral index $\alpha$. We find that the model (eq. \ref{eq:war}) is
most consistent with our measurements for a mean spectral index of
$\alpha=0.7$ (Figure \ref{nordnds}).

\begin{table}
\centering
\caption{$150 \, {\rm MHz}$ differential source counts for FIELD
  I. The columns show bin flux limits, the mean flux density of
  sources in each bin, number of sources, the noise corrected number
  of sources and $dN/dS$ with error.
\label{dnds} }
\begin{tabular}{ccccccc}
\hline
Flux Bin & $S$	& $N$ 	& $N_{c}$ 	& $dN_{c}/dS$  \\
 (mJy)   &(mJy)		&   	&          	& (sr$^{-1}$Jy$^{-1}$)\\
\hline
9.11 --	14.49 &	12.23 &	9 & 32.82 & $(3.03 \pm 0.53) \times 10^6$	\\
14.49 -- 23.06 & 18.48 & 25 & 35.15 & $(2.04 \pm 0.34) \times 10^6$\\
23.06 -- 36.68 & 29.38 & 44 & 44.16 & $(1.61 \pm 0.24) \times 10^6$\\
36.68 -- 58.35 & 44.96 & 38 & 38.00 & $(8.72 \pm 1.41) \times 10^5$ \\
58.35 -- 92.82 & 72.71 & 27 & 27.00 & $(3.89 \pm 0.75) \times 10^5$ \\
92.82 -- 147.67& 118.51 & 23 & 23.00 & $(2.08 \pm 0.43) \times 10^5$ \\
147.67 -- 234.91& 192.87 & 18 &	18.00 & $(1.03 \pm 0.24) \times 10^5$ \\  
234.91 --  373.70 & 306.16 & 10 & 10.00 & $(3.58 \pm 1.13)  \times 10^4$ \\  
373.70 -- 594.48 & 495.95 & 8 &	8.00& $(1.80 \pm 0.64) \times 10^4$ \\ 
594.48 -- 945.71 & 799.70 & 4 &	4.00 &$(5.66 \pm 2.83) \times 10^3$ \\ 
\hline
\end{tabular}
\end{table}

We find that our differential source count (Figure \ref{nordnds})
which is determined over a flux range from $9 \, {\rm mJy}$ to $1 \,
{\rm Jy}$ is well fitted by a single power-law of slope $0.97 \pm
0.07$. We note that \citet{George} have estimated a single power law
of slope 0.72 in the flux range of $20 \, {\rm mJy}$ to $2 \, {\rm
  Jy}$ from a $150 \, {\rm MHz}$ GMRT survey centered around
$\epsilon$ Eridanus. In Figure \ref{nordnds} we show the normalised
differential source count from their survey. We find that they have
under determined the normalised differential source count in the
entire flux range and it is most likely due to the low number of
sources (113) that were used to determine the source count. For
comparison we also plot the combined results from 6C and 7C
survey \footnote{http://web.oapd.inaf.it/rstools/srccnt/150MHz.dat} in
Figure \ref{nordnds}. We find that there is a good agreement between
our source count and those derived from the 6C, 7C survey
\citep{hales,McGilchrist} in the flux range $0.1\, {\rm Jy}$ to $1.0
\, {\rm Jy}$. \citet{intema} have determined the normalised
differential source count at $153 \, {\rm MHz}$ using a catalog of 598
sources in the flux range of $4.75\, {\rm mJy}$ to $3 \, {\rm Jy}$.
We notice that our source counts are roughly equal to their in the
high flux range of $30.0\, {\rm mJy}$ to $1.0 \, {\rm Jy}$. Towards
the lower flux ends ($< 30.0\, {\rm mJy}$) we find their source counts
are increasingly higher compared to our source count. We note that our
estimated source count agrees well with \citet{ishwara10} in the low
flux range ($< 30.0\, {\rm mJy}$) where they found a relatively low
population of sources compared to \citet{intema}.

The simulated source count model proposed by \citet{Jackson} at $151
\, {\rm MHz}$ predicts a flattening in the power law shape below $10
\, {\rm mJy}$ due to a growing population of Fanaroff - Riley
\citep{Fanaroff} class I (FR I) radio galaxies. The turnover in the
source count is a well-known observed feature at $1.4 \, {\rm GHz}$
\citep{condon,Hopkins,Seymour} and occurs around $1 \, {\rm mJy}$,
which is equivalent to $\sim 5 \, {\rm mJy}$ for an average spectral
index of $0.7$. We note that although this flattening is observed in
deep radio surveys at higher frequencies ($1.4 \, {\rm GHz}$,
\citet{Windhorst}; $610 \, {\rm MHz}$, \citealt{Garn,garn}), our
current survey depth is not sufficient to detect this flattening. This
suggests that our catalog is dominated by the classical radio-loud AGN
population which are the predominant sources at higher flux densities.

We note that the predicted angular power spectrum $C_{\ell}$ (Figure
\ref{fig:clfg}) does not change very significantly if we use the
differential source count measured here instead of the foreground
model used for the predictions in Section \ref{maps}.

\section{Diffuse Galactic foreground}
\label{dgf}

\begin{figure*}
\includegraphics[width=55mm,angle=-90]{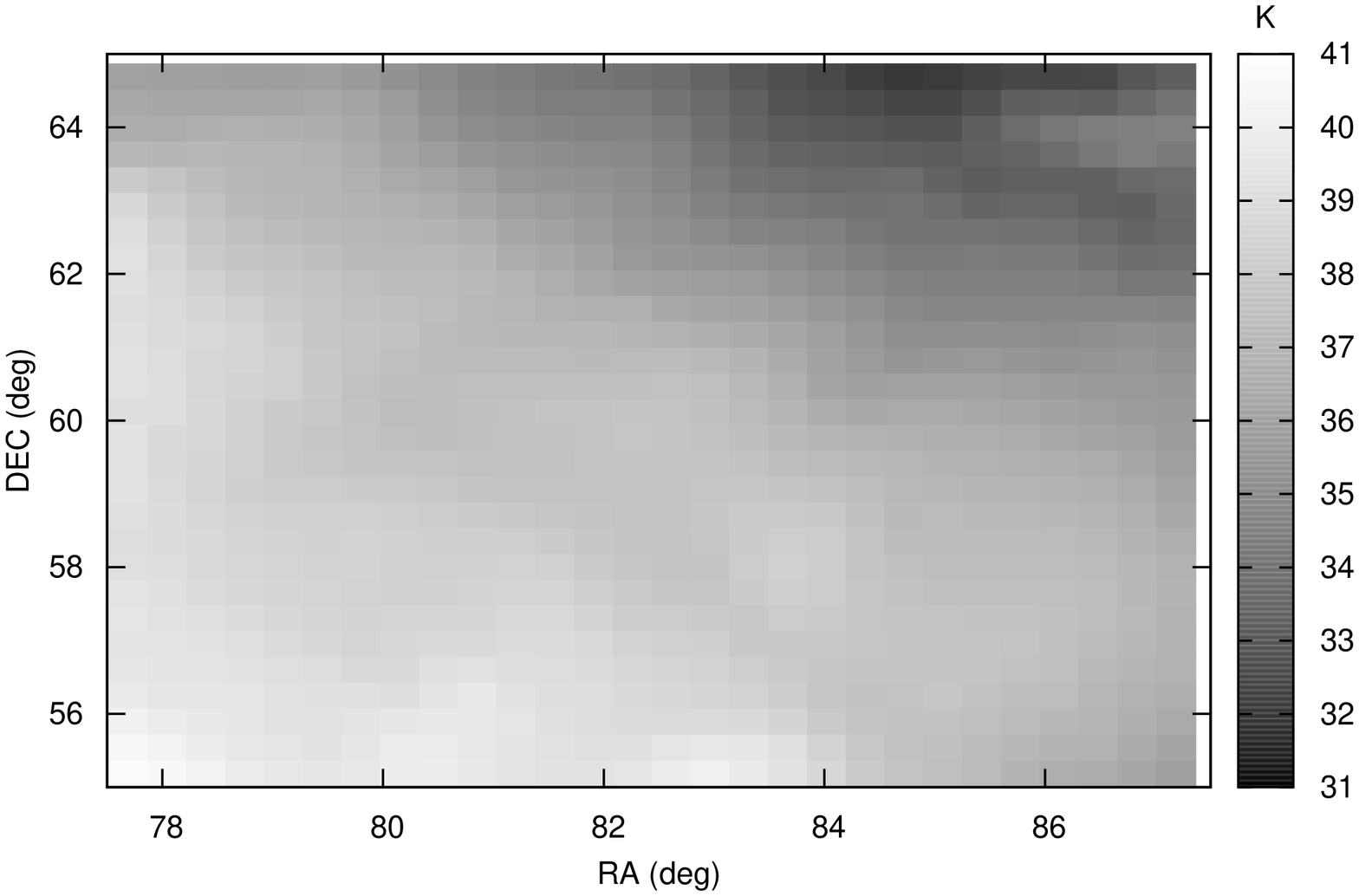}
\includegraphics[width=55mm,angle=-90]{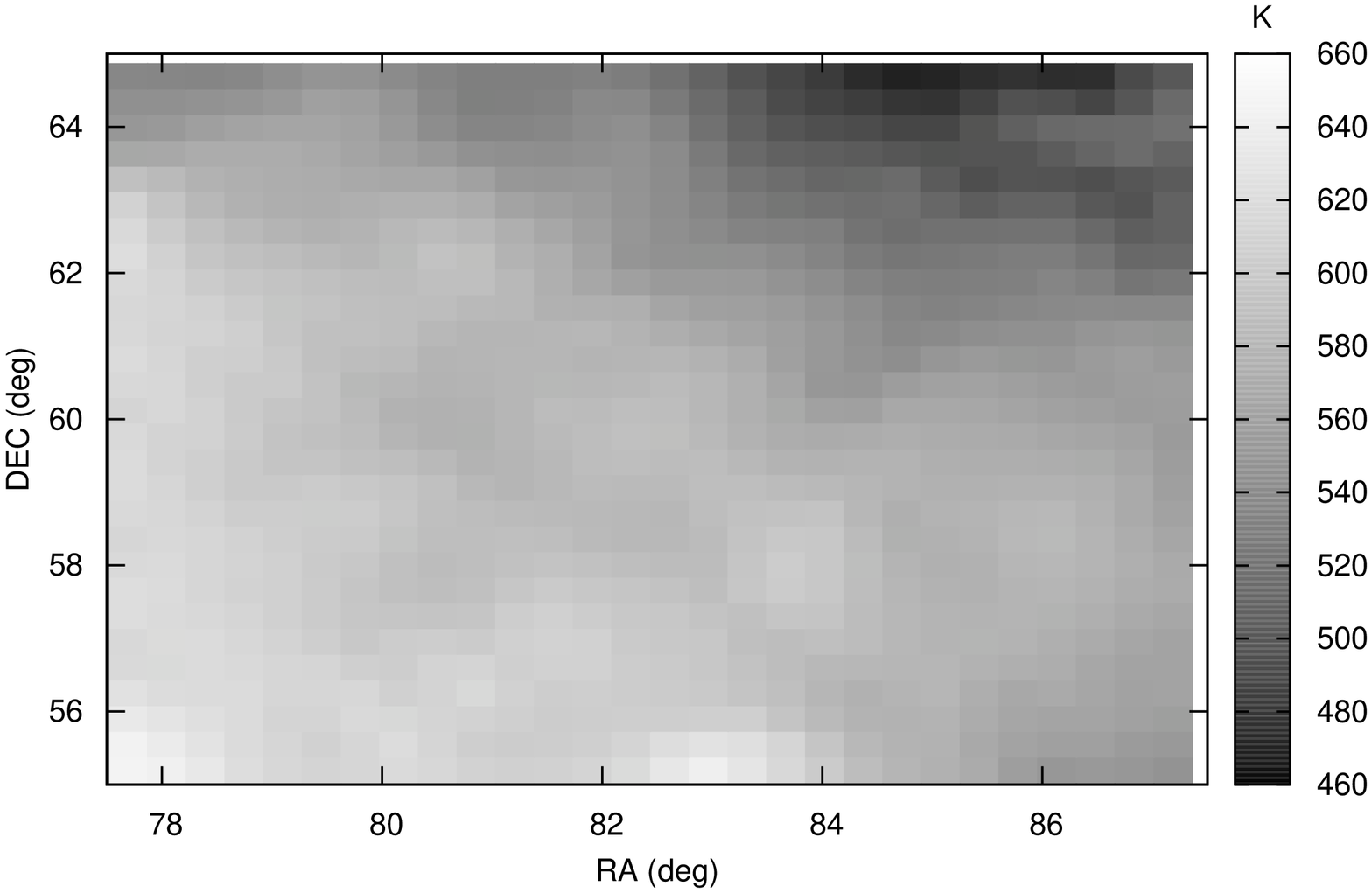}
\caption{The figure in the left panel shows the brightness temperature
  distribution around FIELD I at $408 \, {\rm MHz}$ \citep{haslam}
  after destriping \citep{platania}. In the right panel we show the
  temperature values at $150 \, {\rm MHz}$ where we scale the temperature
  values with the average spectral index map from \citep{platania}. We
  use \citep{lfmap} to generate this figures.}
  \label{fig:has}
\end{figure*}

The diffuse synchrotron radiation from our Galaxy dominates the sky at
low radio frequencies like $150 \, {\rm MHz}$. \citet{haslam} surveyed
the full sky with an angular resolution of $0.85^{\circ}$ at $408 \,
{\rm MHz}$ using single dish observations. The sky brightness
temperature is found to vary in the range $11 \, {\rm K}$ to $4,247 \,
{\rm K}$ across the entire sky. Assuming that the synchrotron
brightness temperature scales as $T \propto \nu^{-2.8}$ with frequency
\citep{platania}, we have a conversion factor of $16.47$ from $408
\,{\rm MHz}$ to $150 \, {\rm MHz}$. This gives the brightness
temperature range from $181 \, {\rm K}$ to $69,948 \, {\rm K}$ across
the entire sky at $150 \, {\rm MHz}$, and $(660,495,330,495) \, {\rm
  K}$ at the respective phases centers of the four fields that we have
observed (Table \ref{tab:obs_sum}). The subsequent analysis is
restricted to FIELD I where we have been able to achieve the best
sensitivity, and point source subtraction is also most
effective. Further this is the lowest Galactic latitude field
($b=13.89^{\circ}$) among our observed fields and it may provide an
upper limit on the expected diffuse foreground emission for four
fields that we have observed. Figure \ref{fig:has} shows the $408 \,
{\rm MHz}$ \citep{haslam} brightness temperature distribution within a
$10^{\circ} \times 10^{\circ}$ region of FIELD I.  The figure also
shows the $150 \, {\rm MHz}$ brightness temperature distribution
obtained by scaling the $408 \, {\rm MHz}$ maps.  The scaling to $150
\, {\rm MHz}$ takes into the angular variation of the spectral index
\citep{platania}, and we have used LFmap \citep{lfmap} to generate the
data for these figures.  The smallest baseline in our observation is
around $U =100$, which corresponds to an angular scale of $U^{-1}
\approx 30'$. Our observations are not sensitive to intensity
variations at angular scales larger than this, and consequently we do
not expect the structures seen in Figure \ref{fig:has} to be imprinted
in our observation. Very little is known about the angular structure
of the diffuse Galactic synchrotron radiation on sub-degree scales.

Model predictions \citep{ali}, which extrapolate the statistical
properties measured at large angular scales and higher frequencies,
predict that point sources are the most dominant contribution at
sub-degree scales.  We expect the diffuse radiation to dominate if the
point sources can be individually modelled and removed with high level
of precession.  \cite{bernardi09} have analyzed $150 \, {\rm MHz}$
WSRT observations where they have subtracted out the point sources to
reveal the structure of the fluctuations in the diffuse Galactic
synchrotron radiation at $13'$ angular scales. We note that their
observation was carried out at a low Galactic latitude ($b=8^{\circ}$)
where the Galactic emission is expected to be relatively larger than
our targeted field. It is expected that we should be able to use the
residual data to characterize the diffuse radiation provided the point
sources have been removed to an adequate level of sensitivity.

We have discussed point source subtraction in Section \ref{ps}.  For
FIELD I, Figure \ref{fig:fields} and Figure \ref{fig:subfields} show
$4.0^{\circ}\times4.0^{\circ}$ continuum images of the field before
and after source subtraction.  All the pixels with flux density
greater than $10 \, {\rm mJy}$/Beam were visually inspected, and those
which appeared to be genuine sources were fitted with Clean Components
using tight boxes. The continuum Clean Components were then subtracted
from the visibility data.  The image, after source subtraction, has
residual flux density in the range of $-14 \, {\rm mJy}$/Beam to $21
\, {\rm mJy}$/Beam arising from imaging artifacts. We recollect here
that these residuals are highly localized in the vicinity of a few
regions that contained the brightest sources (Figure
\ref{fig:subfields}). The bulk of the field is largely free of
artifacts and is consistent with noise.

It is possible that very bright sources beyond the
$4.0^{\circ}\times4.0^{\circ}$ field that we have imaged also
contribute to the measured visibilities.  To account for this
possibility we have also imaged a $8.0^{\circ}\times8.0^{\circ}$
region after the sources within the central
$4.0^{\circ}\times4.0^{\circ}$ region have been subtracted out.  We
have then removed all the sources automatically to a $10 \, {\rm
  mJy}$/Beam level using the $8.0^{\circ}\times8.0^{\circ}$ image.

\begin{figure*}
\includegraphics[width=150mm]{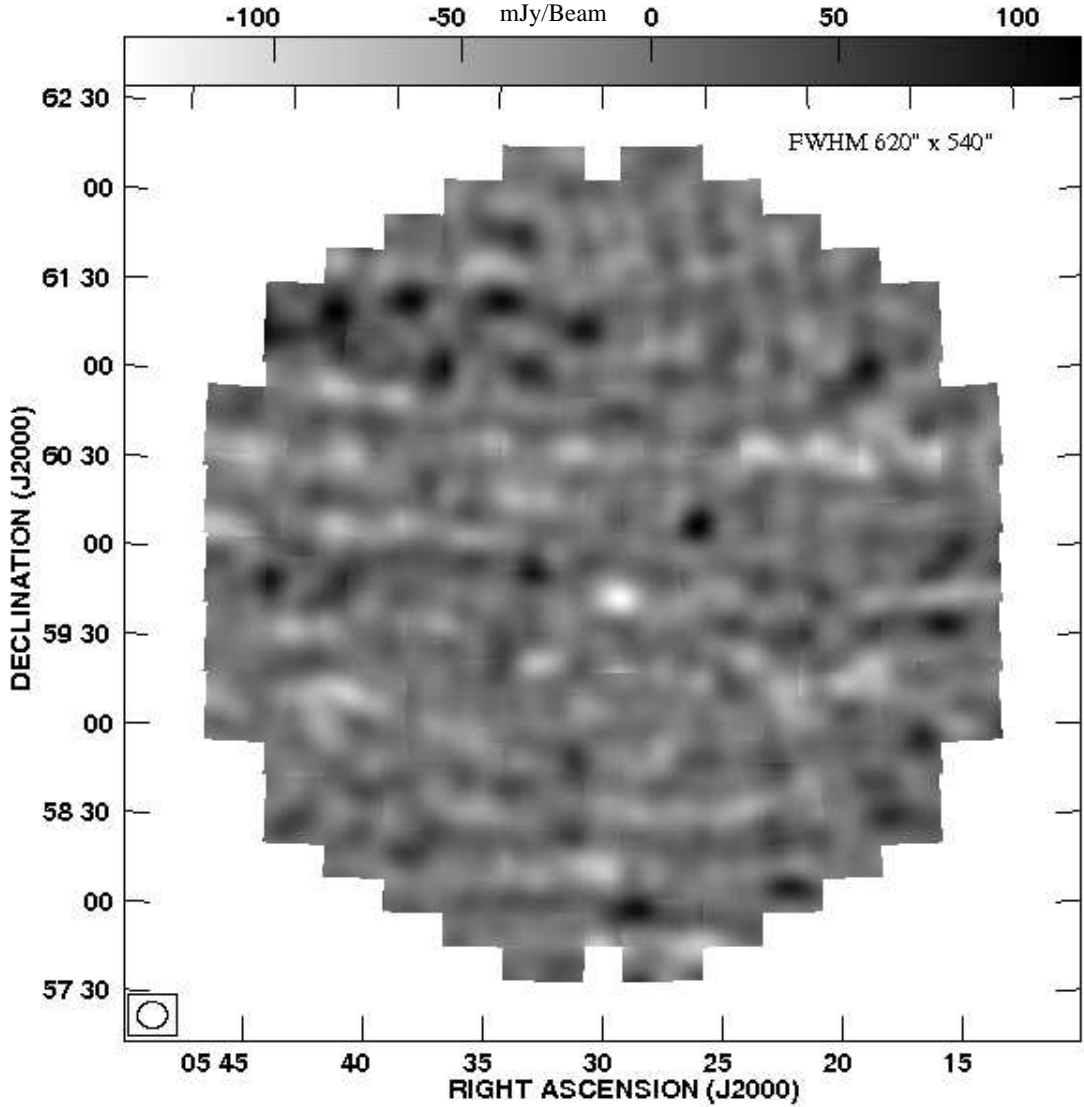}
\caption{This shows the image where we tapered the $uv$ plane at $|\u| =
  200$. The synthesized beam has a FWHM of $620''\times540'' $. All
  the genuine point sources were removed down to $10 \, {\rm mJy}$
  level.  Diffuse structures begin to appear on $> 10^{'}$ scales.
  The brightest structures in this map are at a $\sim 5 \sigma$ level
  compared to the local rms of $23.5 \, {\rm mJy/Beam}$.}
\label{lres2}
\end{figure*}

We expect to see the diffuse background radiation after the point
sources have been removed, however the high resolution
($20''\times18'' $) residual image does not exhibit any diffuse
structure.  The noise and the residuals after point source removal
appear to dominate the high resolution image.  We expect the Galactic
synchrotron radiation to be relatively larger in comparison to the
noise and residuals if we consider larger angular scales. We have
considered the angular scale $10'$, and following \cite{bernardi09} we
have made two images which refer to this angular scale. Initially, we
made an image which does not have any visibilities with baselines
$|\u| < 170$.  This restriction imposes the condition that the
resulting image does not contain any information on angular scales
grater than $10'$.  For the second image (Figure \ref{lres2}), the
visibilities were tapered with a Gaussian in the $uv$ plane so as to
produce a synthesised beam of FWHM $620''\times540''$. This image does
not contain information at angular scales below $10'$.

We see that the image, made by including only the baselines $|\u| >
170$, which does not contain any information above $10'$ looks very
similar to the high resolution image.  Both of these images are
dominated by the noise, and the residuals from the brightest point
sources are seen to be localized near the center of the image. In
contrast, the low resolution image (Figure \ref{lres2}), which does
not contain information at angular scales below $10'$, shows
fluctuations which are uncorrelated with the point source distribution
seen in the high resolution image (Figure \ref{fig:fields}).  The
maximum and minimum values of flux density in Figure \ref{lres2} are
$113 \, {\rm mJy/Beam}$ and $-139 \, {\rm mJy/Beam}$ respectively,
which are comparable to $5\sigma$ where $\sigma$ (rms) is $23.5 \,
{\rm mJy/Beam}$ (Table~\ref{rms}). The individual regions
corresponding to the very high (or low) pixel values also have an
angular extent that is comparable to the synthesized beam.  We
interpret these features, which correspond to brightness temperature
fluctuations of the order of $20 \, {\rm K}$ (Table~\ref{rms}), as a
tentative detection of the Galactic synchrotron radiation at the $10'$
angular scale.

\begin{figure*}
\includegraphics[width=150mm]{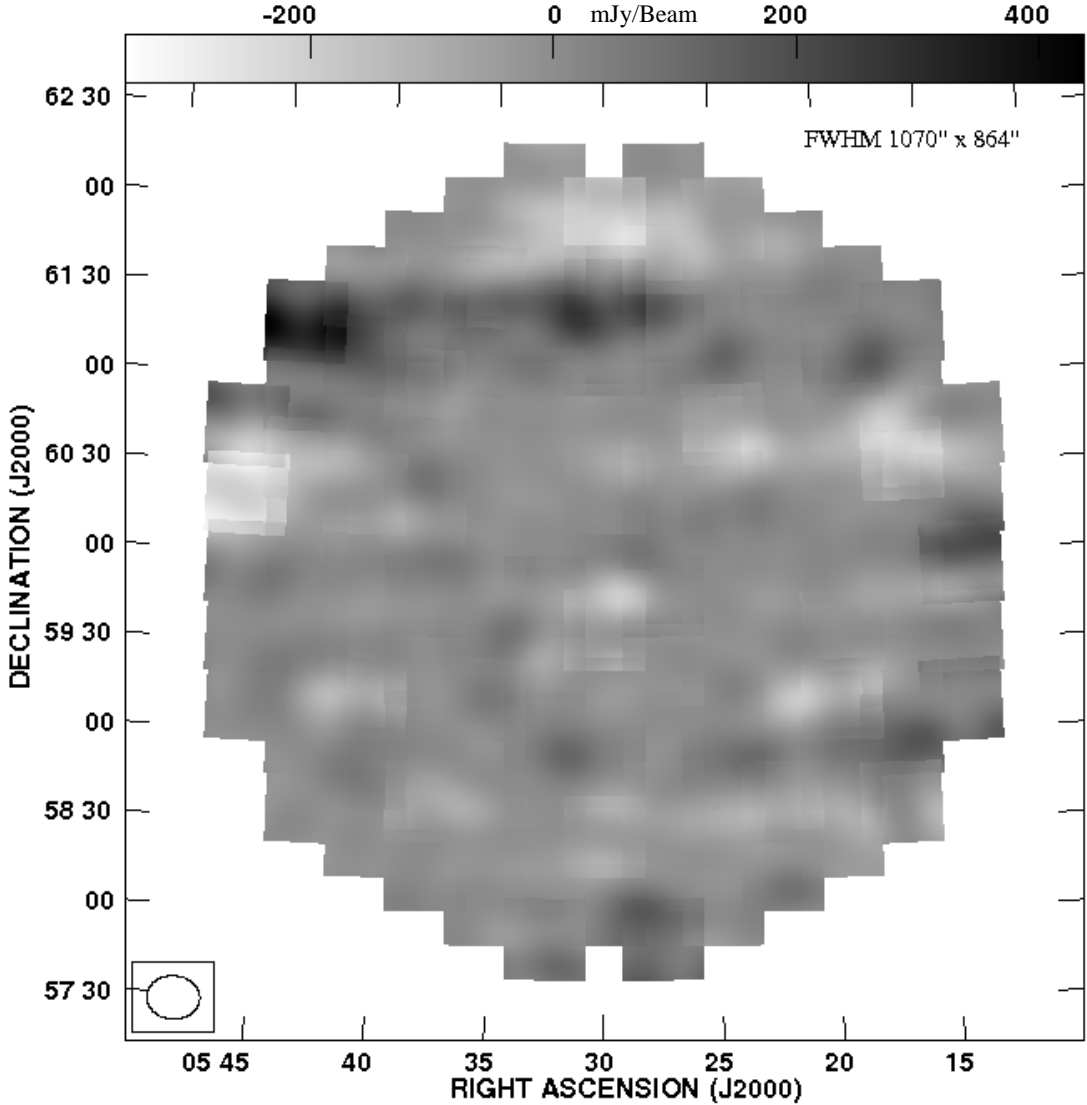}
\caption{This shows the image where we tapered the $uv$ plane at $|\u| =
  100$. The synthesized beam has a FWHM of $1070''\times864''$. All
  the genuine point sources were removed down to $10 \, {\rm mJy}$
  level. The brightest structures in this map are at a $10\sigma$
  level compared to the local rms of $35 \, {\rm mJy/Beam}$.}
\label{lres1}
\end{figure*}

We next increase the $uv$ taper to produce a synthesized beam with a
larger FWHM of $1070''\times 864''$ .  Figure \ref{lres1} shows the
corresponding image which does not have any information at angular
scales below approximately $16'$.  The maximum and minimum values of
flux density in this image are $436 \, {\rm mJy/Beam}$ and $-353\,
{\rm mJy/Beam}$ respectively, which are greater than $10\sigma$ where
$\sigma$ (rms) is $35 \, {\rm mJy/Beam}$ (Table~\ref{rms}). The
individual regions corresponding to the very high (or low) pixel
values have an angular extent that is comparable to, if not bigger
than, the synthesized beam.  In fact the size of these regions
approaches the largest angular scales $(\sim 30')$ that can be probed
in our observation.  We interpret these $10 \sigma$ fluctuations as a
detection of the fluctuations in the Galactic synchrotron radiation at
the $16'$ angular scale.  Converting to brightness temperature
(Table~\ref{rms}), we have a peak fluctuation of $26 \, {\rm K}$ in
our field of view.

Based on our observation we conclude that the residuals after point
source subtraction represent structures in the Galactic synchrotron
radiation on angular scales of $10'$ to $20'$.  The noise and the
residuals from point source subtraction, however, dominate the data at
smaller angular scales where we are unable to make out the Galactic
synchrotron radiation.

\begin{table}
\centering
\caption{Rms fluctuations in the FIELD I as a function of angular resolution 
\label{rms}}
\begin{tabular}{lccc}
\hline
Angular&rms&Conversion factor&rms\\
resolution&(mJy/Beam) & (mJy/Beam) to (K) & (K) \\ 
\hline
$20'' \times 18''$ & 1.30 & 171.80 & 223.34	\\
$620'' \times 540''$ & 23.50 & 0.17 & 4.00	\\
$1070'' \times 864''$ & 35.00 & 0.06 & 2.1	\\
\hline
\end{tabular}
\end{table}

\subsection{Power Spectrum}

\begin{figure}
\includegraphics[width=60mm,angle=-90]{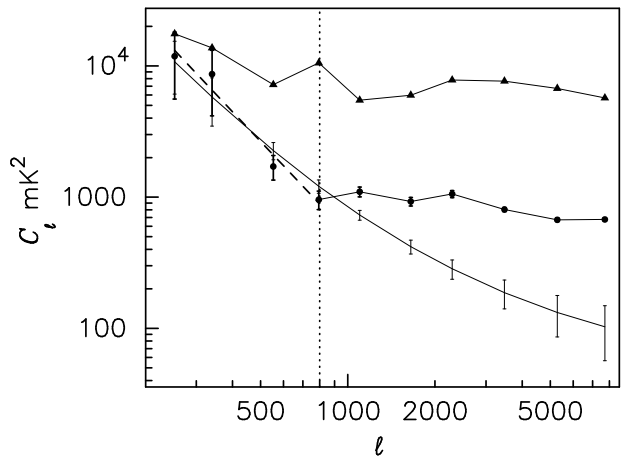}
\caption{ The measured $C_{\ell}$ before (triangle),  and after (circle) 
source subtraction, with   $1\sigma$ error-bars shown only for the
latter. The   dashed line  shows the best fit  power-law  $(\ell \le
800)$ after source removal.  The thin solid   
  line (lowest  curve)   with $1$ - $\sigma$ error-bars shows the
  model prediction of \citealt{ali} assuming that all point sources
  above   $S_{c} = 20 \, {\rm    mJy}$ have been removed. }
\label{diffgalps}
\end{figure}

We have used the visibilities, after point source subtraction, to
estimate the angular power spectrum $C_{\ell} \equiv C_{\ell}(\Delta
\nu=0)$.  The angular power spectrum $C_{\ell}$ (Figure
\ref{diffgalps}) clearly shows two different scaling behavior as a
function of $\ell$.  At low angular multipoles ($\ell \le 800$), which
correspond to angular scales larger than $10'$, we find a steep power
law behavior which is typical of the Galactic synchrotron emission
observed at higher frequencies and larger angular scales
\citep{bennett03}. The angular power spectrum flattens out for $l >
800$, and we find that it remains nearly flat out to $\ell \approx
8,000$ which corresponds to angular scales of $\sim 1^{'}$.  The
nearly flat region arises from the point sources which have a flux
that is below the threshold for source identification and removal, and
hence have not been subtracted from the data. The error's in modelling
and subtracting the identified sources also contribute to this.

It is clear from this analysis that $C_{\ell}$, after point source
subtraction, is dominated by the diffuse Galactic synchrotron emission
at $\ell \le 800$ which corresponds to $\sim 10'$.  We also note that
the angular power spectrum $C_{\ell}$ was directly estimated from the
visibilities using the method discussed in Section \ref{sec3} which
makes absolutely no reference to the image. However, it is reassuring
to note that the measured $C_{\ell}$ is quite consistent with the
behavior that we noticed in the image plane. We do not find any
significant features, other than the imaging artifacts, in the high
resolution residual image and in the residual image made by discarding
the short baselines ($|\u| > 170$). Significant features only appear
in the residual image if the angular resolution is $\ge 10'$. The
measured angular power spectrum predicts the fluctuations in the
Galactic synchrotron emission to be $\sim
\sqrt{\ell(\ell+1)C_{\ell}/2\pi} \simeq 10 \, {\rm K}$ at $\ell=800$
which is comparable to the $20 \, {\rm K}$ features seen in the $10'$
resolution image (Figure \ref{lres2}).  We have used a weighted least
square to fit a power law

\begin{equation}
C_{\ell}=A  \times \left(\frac{1000}{\ell} \right)^{\beta} \,
{\rm   mK^2} 
\label{eq:fit}
\end{equation}

to the measured $C_{\ell}$  for $\ell \le 800$ .  We find the best fit
values   $A=513\pm 41 \, {\rm mK}^2$ and $\beta=2.34 \pm 0.28$ for
which $C_{\ell}$  is  shown in Figure \ref{diffgalps}.  

\cite{bernardi09} have analyzed $150 \, {\rm MHz}$ WSRT observations,
where they have subtracted out point sources above $15 \, {\rm mJy}$
and used the resulting image to estimate the angular power spectrum
$C_{\ell}$. We note that their analysis differs from our in that they
have estimated $C_{\ell}$ from the image whereas we have directly used
the measured visibilities to estimate $C_{\ell}$ (Figure
\ref{diffgalps}).  Our findings, however, are very similar to those of
\cite{bernardi09} who find that the measured $C_{\ell}$ is well
described by a power law for $\ell \le 900$. Their best fit parameters
are $A=253\pm 40 \, {\rm mK}^2$ and $\beta=2.2 \pm 0.3$. The slope, we
note, is consistent with our findings, the amplitude, however, is half
our value.  The difference in amplitude is not surprising as the two
different values refer to observations in two different parts of the
sky.  The amplitude measured by \cite{bernardi09} refers to a field
with Galactic latitude $b = 8^{\circ}$, whereas FIELD I of our
observations is at $b=13.89^{\circ}$. We would generally expect lesser
synchrotron emission in our field which is at a higher elevation from
the Galactic plane, however, the amplitude of $C_{\ell}$ shows the
opposite behaviour.  The variation of the sub-degree scale angular
structure of the Galactic synchrotron emission across different parts
of the sky is not known at present, and it is not possible to make any
conclusive statement about this apparent discrepancy. It is possible
that there may be some extra power coming from the clustering of
unresolved point sources below $\ell \le 800$. We note that our
estimate of the angular power spectrum for the Galactic diffuse
emission (Figure \ref{diffgalps}) matches quite well with the
foreground model prediction of \citet{ali}, assuming that all point
sources above a threshold flux limit of 20 mJy (see for reference top
left panel of Figure \ref{fig:subfields}, FIELD I) have been removed.

Strictly speaking, it is not justified to compare the amplitude of the
angular power spectrum obtained in observations at two different
frequencies unless their slopes (angular spectral index) are
consistent with each other.  We have not found any other observations
(except \citet{bernardi09}) of the northern Galactic plane where the
angular spectral index is consistent with our findings. Despite this,
we have used an earlier work \citep{laporta} at higher frequencies to
estimate the amplitude of the angular power spectrum expected at our
observing frequency.  \citet{laporta} have estimated angular power
spectrum of all-sky total intensity maps at 408 MHz \citep{haslam} and
1420 MHz \citep{R82,R86,R01}. The angular power spectrum of the
Galactic synchrotron emission is measured down to the angular mode of
$\ell=200$ and $\ell=300$ at 408 MHz and 1420 MHz respectively. Using
the best fit parameters (tabulated at $\ell=100$) at 408 MHz and 1420
MHz, we obtain the amplitude of the angular power spectrum
$C_{\ell=200} \simeq 230 \, {\rm mK}^{2}$ and $0.22 \, {\rm mK}^{2}$
respectively close to the Galactic latitude of $b=15^{\circ}$ (close
to FIELD I). By extrapolating the values of $C_{\ell=200}$ from 408
MHz and 1420 MHz using a mean frequency spectral index of $\alpha=2.5$
\citep{decosta} ($C_{\ell} \propto \nu^{-2 \, \alpha}$) we find that
the expected contributions at 150 MHz from the Galactic synchrotron
emission are $3.42 \times 10^4 \, {\rm mK}^{2}$ and $2.08 \times 10^4
\, {\rm mK}^{2}$ respectively. In our observation we find
$C_{\ell=200} = 2.22 \times 10^4 \ {\rm mK}^{2}$ which lies within the
range $2.08 \times 10^4 \, {\rm mK}^{2} \ {\rm to} \ 3.42 \times 10^4
\, {\rm mK}^{2}$, and hence we conclude that our finding is consistent
with the values extrapolated from higher frequency and larger angular
scales.

\citet{giardino01} have analyzed the fluctuations in the Galactic
synchrotron radiation using the $2.3 \, {\rm GHz}$ Rhodes Survey where
they find a slope $\beta=2.43 \pm 0.01$ across the range $2 \le \ell
\le 100$. \citet{giardino02} have analyzed the $2.4 \, {\rm GHz}$
Parkes radio continuum and polarization survey of the southern
Galactic plane where they find a slope $\beta=2.37 \pm 0.21$ across
the range $40 \le \ell \le 250$. Our slope, measured at smaller
angular scales, is consistent with these findings.  There has been
considerable work on modelling the Galactic synchrotron radiation at
the higher frequencies relevant for the Cosmic microwave background
radiation (CMBR) (tens of GHz). These models predict a power law
scaling behaviour $C_{\ell} \propto \ell^{-\beta}$ where $\beta$ have
values in the range $2.4$ to $3$ down to $\ell=900$
\citep{tegmark96,tegmark2000}.

We note that the slope measured by us at $150 \, {\rm MHz}$ is
consistent with these model predictions. \citep{bennett03} have
determined the angular power spectrum of the Galactic synchrotron
radiation using the Wilkinson Microwave Anisotropy Probe (WMAP) data
in the frequency range $23$ to $94 \, {\rm GHz}$. Their measurement is
restricted to $\ell \le 200$ where they find a scaling $C_{\ell} \sim
\ell^{-2}$ which is slightly flatter than the slope we obtain at
smaller angular scales.

\section{Discussion and conclusions}
 
We have analyzed $150 \, {\rm MHz}$ GMRT observations in four
different fields of view (Table~\ref{tab:obs_sum}, Figure
\ref{fig:fields}). We have used the multi-frequency angular power
spectrum (MAPS) $C_{\ell}(\Delta \nu)$ to jointly characterize the
statistical properties of the fluctuations as a function of the
angular multipole $\ell$ and the frequency separation $\Delta\nu$
(Figure \ref{fig:clcont}) across the range $700 \le \ell \le 2 \times
10^4$ and $0 \le \Delta \nu \le 2.5 \, {\rm MHz}$.  We find that the
measured $C_{\ell}$ has values around $10^4 \, {\rm mK^2}$ at $\ell
\sim 1000$, and it drops by around $50 \%$ at $\ell\sim 10^4$. The
measured $C_{\ell}$ is foreground dominated and is more than $7$
orders of magnitude larger than the expected HI signal.

Analytic estimates of the HI signal \citep{BA5} show that we expect
$C^{HI}_{\ell}(\Delta \nu)$ to decorrelate rapidly with increasing
$\Delta \nu$, and fall by $90\%$ or more at $\Delta\nu \ge 0.5 \ {\rm
  MHz}$ . In contrast, the foregrounds from different astrophysical
sources are expected to be correlated over large frequency separations
($\Delta \nu > 2 \, {\rm MHz}$). This holds the promise of allowing us
to separate the signal from the foregrounds \citep{ghosh2}. We find
that the measured $C_{\ell}(\Delta \nu)$ (for reference see Figures
\ref{fig:clnuf1} and \ref{fig:clnuf2}) shows a smooth $\Delta \nu$
dependence with a $20$ to $40 \, \%$ variation across the measured
$2.5 \, {\rm MHz}$ $\Delta \nu$ range. However, contrary to our
expectations, in addition to a smooth behaviour we also notice abrupt
variations and oscillations $(\le 10 \%)$ in the $\Delta \nu$
dependence of the measured $C_{\ell}(\Delta \nu)$. These abrupt
variations and oscillations, whose origin is at present not
understood, poses a severe impediment for foreground removal.
Polynomial fitting along the frequency axis has been extensively
considered for foreground removal.  All attempts in this direction may
fail due to the presence of abrupt variations and small oscillations
which cannot be filtered out with low order polynomials.

The measured $C_{\ell}$'s, we find, are consistent with foreground
models which predict that extragalactic point sources make the most
dominant contribution at the sub-degree scales that we have probed
here (Figure \ref{fig:clfg}).  The brightest source ($\sim 1 \, {\rm
  Jy}$) in each field alone contributes around $10 \%$ of the total
measured $C_{\ell}$.  It is very important to correctly identify the
point sources and subtract these out at a high level of precision. We
have carried out point source subtraction in all the four fields that
we have analyzed here (Figures \ref{fig:bs1} and \ref{fig:subfields}).
Considering only the brightest source, we find that source subtraction
is most effective in FIELD I, where the image also has the lowest rms
noise of $1.3 \, {\rm mJy/Beam}$ (Table~\ref{tab:img_sum}). The
residual artifacts after subtracting out the brightest source, we
find, are within $1.5 \%$ of the of the original source flux ($905 \,
{\rm mJy}$).

The subsequent analysis was restricted to FIELD I which has the lowest
rms noise. The Clean Components corresponding to all the sources above
$10 \, {\rm mJy}$ were subtracted out from the visibility data using
the AIPS task UVSUB. We find that the resulting angular power spectrum
$C_{\ell}$ falls to $\sim 1,000 \, {\rm mK}^2$ in the $\ell$ range
$800$ to $8,000$ (Figure \ref{diffgalps}), which is roughly one-tenth
of the $C_{\ell}$ before source subtraction.  At these multipoles, we
interpret the measured $C_{\ell}$, after source subtraction, as
arising from a combination of the residual artifacts from the bright
sources and the faint sources ($< 10 \, {\rm mJy}$) that have not been
removed.  The behaviour at lower multipoles ($\ell \le 800$), we find,
is well fitted by a power law $C_{\ell}= (513\pm 41 ) \times
(1000/\ell)^{2.34 \pm 0.28} \,{\rm mK^2}$ which we interpret as the
contribution from the diffuse Galactic synchrotron radiation.  The
measured slope is consistent with earlier WSRT $150 \, {\rm MHz}$
observations \citep{bernardi09}, and also with $2.3$ and $2.4 \ {\rm
  GHz}$ results at smaller $\ell$ \citep{giardino01,giardino02},
whereas WMAP finds a flatter slope $C_{\ell} \sim \ell^{-2}$
\citep{bennett03} at smaller $\ell$ and much higher frequencies
($23-94 \ {\rm GHz}$).

\section{Acknowledgment}
 
AG would like to thank M.H.Wieringa for sharing the 3rd chapter of his
thesis. AG thanks Emil Polisensky for a detail discussion on LFmap. AG
acknowledges CSIR, India for providing financial support through
Senior Research Fellowship (SRF). SSA acknowledges the support by DST,
India, under project No.: SR/FTP/PS-088/2010. The data used in this
paper were obtained using GMRT. The GMRT is run by the National Centre
for Radio Astrophysics of the Tata Institute of Fundamental
Research. We thank the GMRT staff for making these observations
possible.


%
%

\appendix
\section{Source catalogue}
\label{catal}
\begin{table*}
\setcounter{table}{0}
\caption{The complete $150 \, {\rm MHz}$ GMRT source catalogue, listed
  in ascending order of right ascension. The full table is available
  online.}
\label{tab:catalogue}
\begin{center}
\begin{tabular}{ccccccccc} \hline
Source Name & RA & Dec & Peak & Local Noise & Int. Flux Density &
Error\\
 & J2000.0 & J2000.0 & mJy beam$^{-1}$ & mJy beam$^{-1}$ & mJy & mJy \\
\hline \hline
GMRTJ051850.9$+$601311.6 & 05:18:50.91 & $+$60:13:11.57 & 43.74 & 2.28 & 66.22 & 5.92 \\  
GMRTJ051931.0$+$601053.1 & 05:19:30.96 & $+$60:10:53.06 & 37.39 & 2.04 & 38.08 & 4.22 \\  
GMRTJ051933.8$+$603114.4 & 05:19:33.75 & $+$60:31:14.40 & 31.42 & 2.38 & 37.90 & 4.31 \\  
GMRTJ051946.6$+$603205.3 & 05:19:46.62 & $+$60:32:05.28 & 125.07 & 2.34 & 206.97 & 6.11 \\  
GMRTJ051958.5$+$600249.9 & 05:19:58.52 & $+$60:02:49.91 & 18.84 & 1.91 & 19.08 & 3.68 \\  
GMRTJ052010.1$+$603703.0 & 05:20:10.06 & $+$60:37:03.02 & 25.16 & 2.22 & 29.63 & 4.86 \\  
GMRTJ052016.0$+$601339.4 & 05:20:16.04 & $+$60:13:39.45 & 31.43 & 1.92 & 42.68 & 4.61 \\  
GMRTJ052052.9$+$600426.4 & 05:20:52.94 & $+$60:04:26.38 & 108.84 & 1.86 & 128.28 & 3.87 \\  
GMRTJ052059.2$+$593306.4 & 05:20:59.24 & $+$59:33:06.42 & 23.36 & 2.02 & 37.92 & 5.07 \\  
GMRTJ052146.0$+$605705.9 & 05:21:46.01 & $+$60:57:05.94 & 47.74 & 2.14 & 68.69 & 5.69 \\  
GMRTJ052147.6$+$591732.7 & 05:21:47.63 & $+$59:17:32.66 & 152.35 & 2.06 & 226.12 & 4.94 \\  
GMRTJ052150.1$+$603309.9 & 05:21:50.06 & $+$60:33:09.86 & 21.56 & 1.90 & 22.62 & 3.67 \\  
GMRTJ052203.1$+$603652.6 & 05:22:03.13 & $+$60:36:52.65 & 30.52 & 1.90 & 34.02 & 3.21 \\  
GMRTJ052207.7$+$594823.6 & 05:22:07.70 & $+$59:48:23.63 & 18.81 & 2.01 & 24.49 & 3.65 \\  
GMRTJ052207.7$+$593820.6 & 05:22:07.71 & $+$59:38:20.57 & 35.74 & 1.92 & 39.56 & 3.17 \\  
GMRTJ052211.4$+$590248.1 & 05:22:11.40 & $+$59:02:48.13 & 30.38 & 2.39 & 33.41 & 4.64 \\  
GMRTJ052212.5$+$602833.9 & 05:22:12.54 & $+$60:28:33.94 & 17.41 & 1.85 & 30.61 & 4.75 \\  
GMRTJ052213.2$+$595938.8 & 05:22:13.21 & $+$59:59:38.81 & 156.82 & 1.82 & 166.62 & 3.16 \\  
GMRTJ052214.0$+$593104.8 & 05:22:13.98 & $+$59:31:04.78 & 19.71 & 1.88 & 48.47 & 6.01 \\  
GMRTJ052214.9$+$590800.9 & 05:22:14.89 & $+$59:08:00.94 & 25.41 & 2.24 & 28.28 & 4.37 \\  
GMRTJ052237.8$+$594709.3 & 05:22:37.79 & $+$59:47:09.29 & 14.87 & 2.00 & 14.99 & 3.00 \\  
GMRTJ052244.9$+$595140.3 & 05:22:44.95 & $+$59:51:40.31 & 17.07 & 1.93 & 37.86 & 4.86 \\  
GMRTJ052246.5$+$591939.2 & 05:22:46.50 & $+$59:19:39.15 & 43.15 & 1.92 & 49.83 & 3.78 \\  
GMRTJ052255.4$+$595129.7 & 05:22:55.44 & $+$59:51:29.70 & 472.06 & 1.93 & 684.06 & 3.60 \\  
GMRTJ052308.4$+$603902.6 & 05:23:08.44 & $+$60:39:02.55 & 73.60 & 1.74 & 84.46 & 3.60 \\  
GMRTJ052311.8$+$600747.8 & 05:23:11.75 & $+$60:07:47.84 & 98.09 & 1.73 & 101.68 & 2.87 \\  
GMRTJ052317.8$+$595149.3 & 05:23:17.84 & $+$59:51:49.26 & 25.85 & 1.92 & 51.16 & 4.27 \\  
GMRTJ052320.8$+$601304.0 & 05:23:20.77 & $+$60:13:03.98 & 492.90 & 1.84 & 552.31 & 3.00 \\  
GMRTJ052321.4$+$594929.7 & 05:23:21.42 & $+$59:49:29.69 & 19.35 & 1.97 & 32.68 & 3.84 \\  
GMRTJ052324.6$+$594703.6 & 05:23:24.62 & $+$59:47:03.62 & 275.72 & 1.97 & 566.23 & 4.39 \\  
GMRTJ052326.4$+$592542.5 & 05:23:26.41 & $+$59:25:42.55 & 15.58 & 1.75 & 17.61 & 3.00 \\  
GMRTJ052326.8$+$590426.4 & 05:23:26.80 & $+$59:04:26.43 & 109.28 & 2.05 & 110.68 & 3.88 \\  
GMRTJ052337.4$+$591059.7 & 05:23:37.40 & $+$59:10:59.68 & 53.94 & 1.95 & 59.38 & 3.35 \\  
GMRTJ052337.8$+$600018.7 & 05:23:37.83 & $+$60:00:18.68 & 23.14 & 1.66 & 24.24 & 2.62 \\  
GMRTJ052356.3$+$603833.7 & 05:23:56.29 & $+$60:38:33.69 & 60.73 & 1.65 & 110.48 & 4.52 \\  
GMRTJ052402.5$+$593800.5 & 05:24:02.55 & $+$59:38:00.52 & 15.94 & 1.63 & 46.61 & 5.60 \\  
GMRTJ052408.9$+$585659.4 & 05:24:08.87 & $+$58:56:59.40 & 205.90 & 2.12 & 225.58 & 4.24 \\  
GMRTJ052412.9$+$594652.1 & 05:24:12.91 & $+$59:46:52.06 & 296.69 & 1.77 & 338.91 & 2.83 \\  
GMRTJ052414.2$+$600001.3 & 05:24:14.21 & $+$60:00:01.26 & 79.39 & 1.62 & 83.19 & 2.64 \\  
GMRTJ052416.1$+$601830.4 & 05:24:16.10 & $+$60:18:30.44 & 14.86 & 1.75 & 15.66 & 2.75 \\  
GMRTJ052417.4$+$602737.2 & 05:24:17.39 & $+$60:27:37.23 & 46.03 & 1.57 & 52.44 & 3.01 \\  
GMRTJ052418.4$+$602708.0 & 05:24:18.37 & $+$60:27:08.00 & 55.73 & 1.57 & 57.65 & 2.68 \\  
GMRTJ052424.9$+$592200.1 & 05:24:24.90 & $+$59:22:00.14 & 16.30 & 1.70 & 18.10 & 2.76 \\  
GMRTJ052426.9$+$595234.0 & 05:24:26.94 & $+$59:52:34.00 & 21.92 & 1.69 & 24.71 & 2.73 \\  
GMRTJ052433.3$+$591759.1 & 05:24:33.28 & $+$59:17:59.13 & 23.50 & 1.75 & 34.46 & 3.86 \\  
GMRTJ052436.8$+$610912.2 & 05:24:36.78 & $+$61:09:12.23 & 32.26 & 1.99 & 36.01 & 4.55 \\  
GMRTJ052440.6$+$585854.4 & 05:24:40.56 & $+$58:58:54.44 & 126.28 & 2.07 & 146.51 & 4.12 \\  
GMRTJ052448.1$+$602715.1 & 05:24:48.11 & $+$60:27:15.11 & 13.49 & 1.56 & 14.77 & 2.60 \\  
GMRTJ052452.7$+$595004.5 & 05:24:52.67 & $+$59:50:04.49 & 16.56 & 1.67 & 21.46 & 2.90 \\  
GMRTJ052452.9$+$605641.9 & 05:24:52.91 & $+$60:56:41.90 & 30.12 & 1.78 & 31.98 & 3.26 \\  
GMRTJ052456.7$+$584255.2 & 05:24:56.75 & $+$58:42:55.18 & 42.00 & 2.42 & 43.78 & 4.84 \\  
GMRTJ052506.4$+$600128.6 & 05:25:06.39 & $+$60:01:28.62 & 14.41 & 1.53 & 15.20 & 2.29 \\  
GMRTJ052529.0$+$605027.3 & 05:25:29.01 & $+$60:50:27.33 & 56.66 & 1.62 & 75.25 & 3.75 \\  
GMRTJ052533.3$+$603155.1 & 05:25:33.27 & $+$60:31:55.06 & 134.08 & 1.55 & 176.86 & 3.13 \\  
GMRTJ052536.5$+$594516.9 & 05:25:36.48 & $+$59:45:16.90 & 11.05 & 1.68 & 12.30 & 2.31 \\  
GMRTJ052541.5$+$601245.2 & 05:25:41.46 & $+$60:12:45.20 & 14.55 & 1.70 & 34.63 & 4.16 \\  
GMRTJ052548.4$+$591802.6 & 05:25:48.41 & $+$59:18:02.63 & 67.43 & 1.66 & 72.51 & 2.96 \\  
GMRTJ052600.1$+$603314.9 & 05:26:00.10 & $+$60:33:14.93 & 13.35 & 1.52 & 16.32 & 2.94 \\  
GMRTJ052645.5$+$583921.0 & 05:26:45.51 & $+$58:39:20.99 & 80.73 & 2.36 & 83.13 & 4.53 \\  
GMRTJ052647.5$+$601450.0 & 05:26:47.48 & $+$60:14:49.97 & 47.41 & 1.72 & 200.70 & 6.16 \\  
\hline
\end{tabular}
\end{center}
\end{table*}

\end{document}